\documentclass[aps,pra,reprint,amsmath, amssymb,superscriptaddress,showpacs]{revtex4-2}

\usepackage[colorlinks = true, linkcolor = blue, urlcolor  = blue, citecolor = blue, anchorcolor = blue]{hyperref}
\usepackage{xspace}
\newcommand*{\ket}[1]{\ensuremath{\left|#1\right>}\xspace}%
\usepackage{multirow, amsmath, array, booktabs, color, bm}
\usepackage{longtable,mathrsfs}
\usepackage{graphicx, epsfig, latexsym, amssymb}
\usepackage[section]{placeins}
\usepackage[textsize=footnotesize]{todonotes}
\usepackage{dcolumn}

\begin{document}
\title{Tunable ultrahigh reflection with broadband via collective atom-atom interaction in waveguide-QED system}
\author{Xin Wang}
\affiliation{School of Physics, Sun Yat-sen University, Guangzhou 510275, China}
\author{Junjun He}
\affiliation{School of Physics, Sun Yat-sen University, Guangzhou 510275, China}
\author{Zeyang Liao}
\email[E-mail:]{liaozy7@mail.sysu.edu.cn}
\affiliation{School of Physics, Sun Yat-sen University, Guangzhou 510275, China}
\author{M. Suhail Zubairy}
\affiliation{Institute for Quantum Science and Engineering (IQSE) and Department of Physics and Astronomy, Texas A$\&$M University, College Station, Texas 77843-4242, USA}
 

\begin{abstract}

        We present a scheme for achieving broadband complete reflection by constructing photonic bandgap via collective atom-atom interaction in a one-dimensional (1D) waveguide quantum electrodynamics (QED) system. Moreover, we propose several strategies to further expand the ultrahigh reflection windows, including increasing the number of atoms with separations near the Bragg distance and inducing gradient frequency modulation among the atoms. The center frequency and bandwidth of the ultrahigh reflection window are dynamically adjustable by applying external electromagnetic field. The results here can enrich the many-body physics of waveguide-QED system and offer a pathway for achieving broadened ultrahigh reflection in a controllable way, which can find important applications in the realms of chip-integrated band filter, quantum storage, optical switching, and wavelength-selective devices.
        

\end{abstract}
\maketitle


\section{\label{sec:intro}Introduction}

Engineering the interaction between quantum emitters and their electromagnetic environment serves as a fundamental building block of contemporary quantum science and technology \cite{raimond2001,kimble2008,Gonzalez-Tudela2024}. However, the interaction between an atom and a single photon in free space is typically weak due to the large mismatch between the photon modes and the atomic emission. By focusing the light into a diffraction-limited spot, a single photon can be reflected by a single atom with a probability close to $10\%$ \cite{Tey2008}. When multiple atoms interact with a shared electromagnetic environment, collective effects can enhance the light-atom interaction and increase the photon reflectivity \cite{dicke1954, Rui2020}. In contrast to the free-space scenario, if an atom strongly couples to a single-mode cavity or waveguide, it can effectively emit most of its energy into the cavity or waveguide mode \cite{wilk2007,arcari2014,Scarpelli2019}. Consequently, a single photon resonant with the atomic transition frequency can be almost completely reflected in a waveguide-QED system \cite{shen2005,Astafiev2010,chen2011}. However, for a single atom, this complete reflection only occurs at the exact resonant frequency,  and rapidly diminishes as the frequency deviates from resonance.

In waveguide-QED systems, the atom-atom interactions mediated by the waveguide mode are inherently long-range, offering a platform to explore exotic quantum many-body effects.\cite{zheng2010,douglas2015,Cheng2017,poshakinskiy2021,r.holzinger2022,silviacardenas-lopez2023,fayard2021,Xing2022,sheremet2023,Xing2024a}. For Bragg-spaced atomic arrays, where the interatomic distance is $ d=m\lambda_0/2 $ (with $ m=1,2,\cdots $), collective interactions can give rise to superradiant states with decay rates of $N\Gamma_{0}$, effectively transforming the atom array into a highly reflective mirror \cite{Deutsch1995,t.s.tsoi2008,Schilke2011,liao2015,corzo2016,sorensen2016}. The superradiant effect in Bragg-spaced arrays can be harnessed for the deterministic generation of multiple-photon states \cite{gonzalez-tudela2015,gonzalez-tudela2017,xing2024}, while subradiant states hold potential for quantum computation based on decoherence-free subspaces \cite{Paulisch2016} and ultrahigh precision quantum sensing \cite{Liao2017OE}. Moreover, two Bragg-spaced atomic arrays can form an effective cavity, facilitating the exploration of cavity quantum electrodynamics \cite{Zhou2008,chang2012,Guimond2016,mirhosseini2019,Zhou2022,Zhou2023} and the generation of single-photon frequency comb \cite{liao2016a}. Despite the substantial reflectivity of Bragg-spaced atom arrays, ultrahigh reflection (i.e., reflectivity larger than $99\%$) is restricted to a small region near the resonant component. This raises the intriguing question of whether the ultrahigh reflection can be achieved over a broad frequency range or not.

Achieving total reflection across a broad bandwidth has been explored through methods such as inducing anomalous dispersion in atomic vapors \cite{wicht1997,g.s.pati2007,yiqiuma2015,othman2018}, stimulated Brillouin scattering \cite{yum2013a}, or utilizing specially designed metamaterials \cite{shi2013,moitra2014}. 
Here, we present an alternative method to attaining ultrahigh reflection with broad bandwidth via collective atom-atom interaction in waveguide-QED systems. This phenomenon can be elucidated by the formation of an energy bandgap induced by collective many-body interactions. We further propose two strategies to broaden the ultrahigh reflection bandwidth: increasing the number of atoms with separations near the Bragg distance and inducing gradient frequency modulation among the atoms. 
The center frequency and bandwidth of these broadened ultrahigh reflection spectra are dynamically adjustable by applying external electromagnetic field which is usually not convenient for the traditional dielectric Bragg reflectors \cite{Bendickson1996}. The results found here hold significant potential for applications in broadband quantum memory \cite{Erhan2011,Guo2019,moiseev2021,Arnold2022}, optical switching \cite{xu2012,Xia2013,Zhu2024}, band-rejection filters~\cite{vengsarkar1996long,Lin2023}, 
wavelength-selective devices~\cite{baumann1996compact,Chen2021}, and wavelength division multiplexing~\cite{breglio2006,Lin2024}.


\section{\label{sec:theory}Collective reflection in waveguide-QED system}

\begin{figure}
    \centering
    \includegraphics[width=0.48\textwidth]{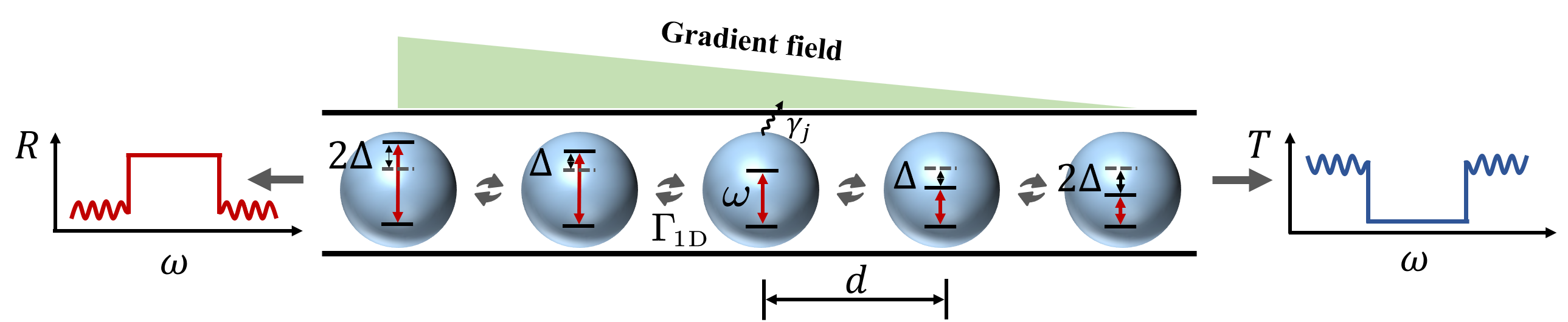}
    \caption{A single photon is scattering by an atomic array coupled to a 1D waveguide, where the transition frequencies of each atom can be tuned by external electromagnetic field on demand.}
    \label{fig:model}
\end{figure}

The schematic model is depicted in Fig.~\ref{fig:model}, where $ N $ two-level atoms separated by an interval $ d $ couple to a 1D single-mode waveguide and the transition frequencies of each atom can be tuned by external electromagnetic field on demand. The Hamiltonian is expressed as \cite{liao2016c,liao2020} 
\begin{equation} 
    \begin{aligned} 
        H = & \hbar \sum_{j=1}^{N} \left(\omega_j-i \frac{\gamma_j}{2}\right) \hat{\sigma}_j^{+} \hat{\sigma}_j^{-}+\hbar \sum_k \omega_k \hat{a}_k^{\dagger} \hat{a}_k \\ 
            & +\hbar \sum_{j=1}^{N} \sum_k\left(g_k \mathrm{e}^{i k z_j} \hat{a}_k \hat{\sigma}_j^{+}+g_k^* \mathrm{e}^{-i k z_j} \hat{a}_k^{\dagger} \hat{\sigma}_j^{-}\right), 
    \end{aligned}
    \label{eq:Hamiltonian}
\end{equation}
where the first term is atomic Hamiltonian with $ \omega_{j} $ and  $ \gamma_{j} $ being the transition frequency and external dissipation rate of the  $ j $th atom positioned at $ z_j $, and $ \sigma_j^{+} = |e\rangle_j\langle g| $ ($ \sigma_j^{-}=|g\rangle_j\langle e| $) being the raising (lowering) operator. 
The second term describes the energy of waveguide photons, where $ \hat{a}_k^{\dagger} (\hat{a}_{k}) $ is the creation (annihilation) operator with wave vector $ k $ and angular frequency $ \omega_k $. 
The last term encapsulates the interaction between the guided photons and the atoms with coupling strength $ g_{k} $.

\begin{figure}
    \centering
    \includegraphics[width=1\columnwidth]{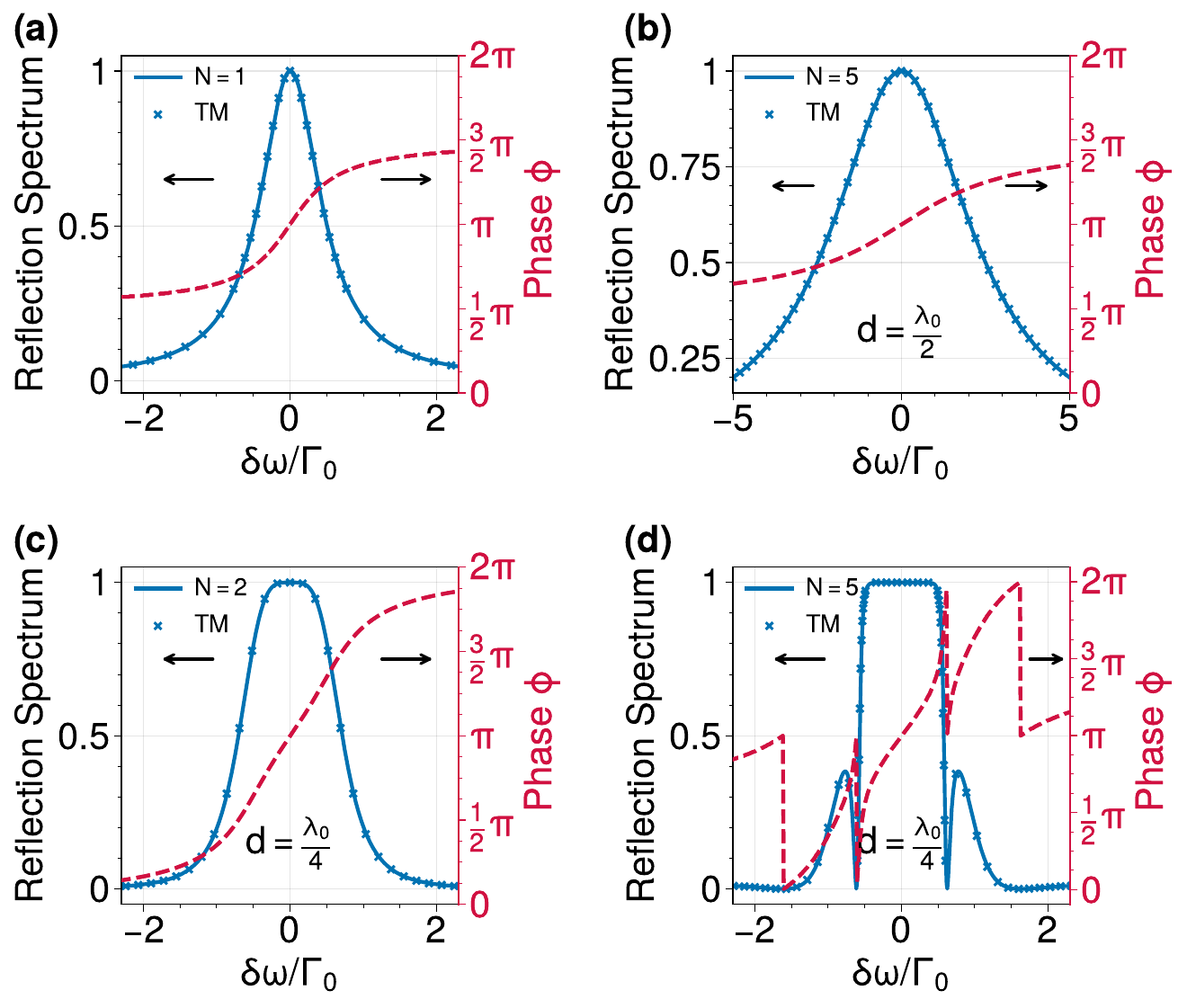}
    \caption{Reflection spectra (blue curves) and reflection phases (red curves) when (a) $ N = 1 $, (b) $ N = 5 $ at Bragg condition ($ d=\frac{\lambda_0}{2} $ ), (c) and (d) at anti-Bragg condition ($ d=\frac{\lambda_0}{4} $). The blue 'x' represents the result calculated by transfer matrix formalism. 
}
    \label{fig:reflectivity}
\end{figure}

Consider that a single photon with wave vector $ k $ is injected from the left end of the waveguide, while all the atoms are in the ground state initially.  
After scattering, the reflection amplitude is given by~\cite{liao2015,liao2016c}
\begin{equation} 
    \begin{aligned}
        r(\omega)=-\frac{1}{2}\sum_{j=1}^{N}\sum_{l=1}^{N}\sqrt{\Gamma_{j}\Gamma_{l}}\mathrm{e}^{ik(z_j+z_l)} M_{jl}^{-1}(\omega)
    \end{aligned}
    \,,
    \label{eq:r_quantum}
\end{equation}
where $ \Gamma_{j} $ is the decay rate of the $ j $th atom into the waveguide mode. The matrix $ \mathbf{M}(\omega) $ is an $ N \times N $ matrix with matrix element $ M_{jl}(\omega)=(\sqrt{\Gamma_{j}\Gamma_{l}}/2)e^{ik|z_j-z_l|}+[\gamma_j/2-i(\omega-\omega_j)]\delta_{jl} $, and $ \delta_{jl} $ represents the Kronecker delta function. The reflection amplitude can also be expressed as $ r(\omega)=\sqrt{R(\omega)}e^{i\phi(\omega)} $, where $ R(\omega)=|r(\omega)|^2 $ denotes the reflectivity and $ \phi(\omega) $ represents the reflection phase. For a single atom without external dissipation ($ \gamma = 0 $), the reflection amplitude is given by $r_{j}(\omega) =-\Gamma_{j}/(\Gamma_{j}-2i\delta\omega )$ \cite{liao2015} which has a typical Lorentzian line shape with a full width at half maximum (FWHM) of $ \Gamma_{j} $.  The photon is completely reflected on resonance, accompanied by a $ \pi $ phase shift (Fig.~\ref{fig:reflectivity}~(a)). In contrast, frequencies detuned from resonance exhibit only partial reflection. However, with multiple atoms, the reflection spectrum undergoes significant changes due to collective atom-atom interactions.

For multiple identical atoms with $\omega_j=\omega_0$ and $\Gamma_{j} \equiv \Gamma_{\text{1D}}=\Gamma_{0}$, by tracing out the photon degree of freedom, the effective Hamiltonian of the atomic system is~\cite{asenjo-garcia2017,albrecht2019}
\begin{equation} \label{eq:effectiveh}
    H_{\text{eff}}=\omega_{0}\sum_{i=1}^{N}\hat{\sigma}_i^{+} \hat{\sigma}_i^{-}-i \frac{\Gamma_{0}}{2} \sum_{j,l=1}^{N} \mathrm{e}^{i k |z_{j}-z_{l}|} \hat{\sigma}_j^{+} \hat{\sigma}_l^{-} 
    \,.
\end{equation}
The reflection amplitude in Eq.~\eqref{eq:r_quantum} can be rewritten as (see Subsec.~\ref{app:subsec_eq-4} in Appendix or \cite{PhysRevA.95.033818})
\begin{equation}
r(\omega)=-i \frac{\Gamma_{0}}{2} \sum_{\xi=1}^{N}\frac{(\vec{\psi}_{in}^{T}\cdot\vec{v}_{\xi})^{2}}{\omega-\lambda_{\xi}}
\end{equation}
where $\vec{\psi}_{in}^{T}=(\mathrm{e}^{ikz_1},\mathrm{e}^{ikz_2},\cdots,\mathrm{e}^{ikz_N})$, and $\lambda_{\xi}$ $(\vec{v}_{\xi})$ are eigenvalues (eigenvectors) of $H_\mathrm{eff}$. It is evident that the overall reflection amplitude results from the interference of different eigenchannels.

\subsection{Atom array with Bragg spacing}
First, we consider the atromic array configured with Bragg spacing, i.e., $ d=m\lambda_0/2 $ where $ m $ is a positive integer. Under this condition, only one eigenstate of the effective Hamiltonian in the single-excitation subspace is superradiant with an enhanced decay rate of $N\Gamma_{0}$, while the remaining eigenstates are dark and decoupled from the waveguide.  Consequently, the atomic array behaves analogously to the single-atom case but with a significantly amplified decay rate of $N\Gamma_{0}$. An illustrative example is presented in Fig.~\ref{fig:reflectivity}~(b) where $N=5$ and $ d=\frac{\lambda_0}{2} $. Compared with the single-atom scenario depicted in Fig.~\ref{fig:reflectivity}~(a), where the reflection spectrum exhibits a full width at half maximum (FWHM) of $ \Gamma_0 $, the spectrum for the 5-atom system has obviously a broaden reflection spectrum with the FWHM being $ 5 \Gamma_0 $. The phase shifts of the reflection in both cases are similar to each other indicating by the red dashed curves in Figs. ~\ref{fig:reflectivity}~(a) and ~\ref{fig:reflectivity}~(b). However, since there is only one reflection channel in the atom array with Bragg spacing, perfect reflection can only appear at the resonant frequency. 

\subsection{\label{subsec:anti_Bragg}Atom array with anti-Bragg spacing}

Then, we consider the scenario of anti-Bragg condition where the atomic separation is $ d=(2m+1)\lambda_0/4 $, $ m=1,2,\cdots $. The reflection in this case is thought to be minimized due to destructive interference among waves reflected by different atoms. 
The reflection spectra when $N=2$ and $N=5$ with anti-Bragg spacing (e.g., $d=\lambda_{0}/4$) are shown in Figs. ~\ref{fig:reflectivity}~(c) and ~\ref{fig:reflectivity}~(d), respectively. Different from the Lorentzian lineshape in the Bragg spacing, the reflection profile for the anti-Bragg spacing exhibits a finite ultrahigh reflection frequency range centered around the resonance, as illustrated in the blue solid curves in Figs. ~\ref{fig:reflectivity}~(c) and ~\ref{fig:reflectivity}~(d). Comparing the results with $N=2$ and $N=5$, we can see that the ultrahigh reflection frequency window is wider and the slope of the edge is sharper when the number of atoms increases. In addition, the reflection phase $ \phi(k) $ when $N=2$ smoothly changes from $0$ to $2\pi$ when sweeping the incident frequency (see red dahsed curve in Fig. ~\ref{fig:reflectivity}~(c)). In contrast, the reflection phase $ \phi(k) $ when $N=5$ exhibits several abrupt $ \pi $ phase transitions at specific frequencies. 
It is noteworthy that the reflectivity vanishes at frequencies associated with these abrupt phase changes, indicating that zero reflection arises from destructive interference. 
For $ N $ atoms, there are $ N-1 $ frequencies where these sudden $ \pi $ phase shifts occur, resulting in $ N-1 $ points of zero reflection.

This broadened ultrahigh reflection window bears resemblance to the photonic stop band observed in distributed Bragg reflectors (DBRs)~\cite{Bendickson1996,sheppard1995,jiang2022a}. 
Thereby, we employ the classical multiple scattering theory and transfer matrix (TM) theory, typically utilized in Bragg dielectric systems, alongside the single-atom reflectivity and transmittivity, to compute the reflectivity of the entire system (see Subsec.~\ref{app:subsec-B} in Appendix). The results, depicted as blue 'x' in Fig.~\ref{fig:reflectivity}, perfectly overlap with those obtained from the full quantum theory, confirming that both theories are fundamentally equivalent in describing the reflection in the single-photon excitation regime in waveguide-QED systems.
Nevertheless, it is crucial to note that beyond this regime, the two theories may diverge due to the emergence of nonlinear effects, such as saturation and strong photon-photon interactions \cite{shen2007,roy2017,liao2020,sheremet2023}.

%
%
%

%

\begin{figure*}
    \centering
    \includegraphics[width=0.7\linewidth]{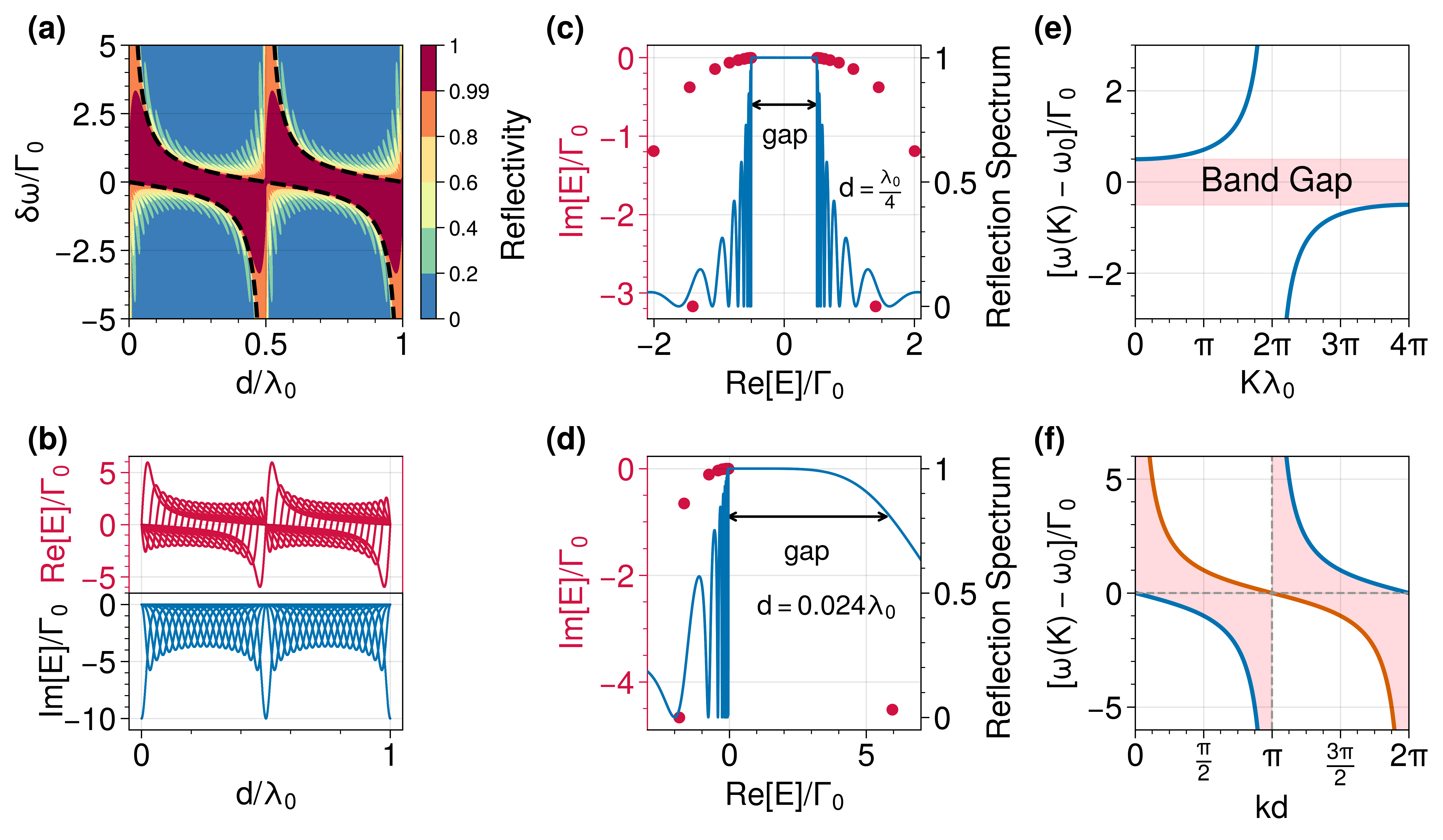}
    \caption{        
        (a) Reflection spectra with varying atom separations. The dark red area marks the ultrahigh reflection region with $R\leq 0.99$. The black dashed curve marks the energy band gap for infinite long atom chain. (b) Real and imaginary part of eigenvalues of the effective Hamiltonian. The Correspondence between the eigenvalues and the reflection spectra: (c) $ d = 0.25 \lambda_0 $ (d) $ d = 0.024 \lambda_0 $. $N=20$ and $\gamma_{j}=0$ for all four subfigures. (e) Dispersion relation for an infinite identical atomic chain when $ d = 0.25 \lambda_0 $. (f) The width of the band gap varies with $k d$.
}
    \label{fig:Heff}
\end{figure*} 

\subsection{Flat-band ultrahigh reflection window and its mechanism}

Next, we highlight a fact that across nearly all atomic separation values, except those satisfying the Bragg condition, the corresponding refection spectra exhibit ultrahigh reflectivity with specific bandwidth ranges. 
For a twenty-atom configuration presented in Fig.~\ref{fig:Heff}~(a), we illustrate the reflection spectra for a photon scattering by atomic arrays with varying atomic separations. 
The dark-red region signifies the ultrahigh reflection band where the reflectivity exceeds $ 99 \% $. When $ d =\lambda_0 /2$, only the resonant frequency is completely reflected. However, for other separations, there is always a finite bandwidth of ultrahigh reflection window. Specifically, when $ d = \lambda_0 / 4 $, the ultrahigh reflection window is symmetric around the resonant frequency, while it becomes asymmetric for other values of $ d $. Thus, by tuning the transition frequency of the atoms or the effective atom-atom separation, we can shift the overall ultrahigh reflection window.

To elucidate the underlying mechanisms of the aforementioned reflection spectra, we delve into the intrinsic characteristics of the non-Hermitian effective Hamiltonian (Eq.~\eqref{eq:effectiveh}) whose eigenvalues are complex. 
The real (Re$ [E] $) and the imaginary (Im$ [E] $) components of these eigenvalues correspond to the collective energy shifts and collective dissipation rates of the system, respectively.
Still considering the $ N = 20 $ case, Fig.~\ref{fig:Heff}~(b) illustrates the real (red solid lines) and imaginary (blue solid lines) eigenvalues within the single-excitation subspace as a function of varying atomic separations.
Due to the collective interactions among the atoms, the energy is split into twenty sublevels, each exhibiting distinct decay rates. Some of these eigenstates exhibit superradiant behavior (-Im$ [E]>\Gamma_{0}/2 $), while others are subradiant (-Im$ [E]<\Gamma_{0}/2 $). 
One can see that the region enclosed by these eigenenergies closely resembles the reflection spectra illustrated in Fig.~\ref{fig:Heff}~(a), indicating that the ultrahigh reflection arises from constructive interference from the reflections of these eigensublevels, as described in Eq.~\eqref{eq:effectiveh}. 
%

%
To further clarify the formation of the ultrahigh reflection window, we plot the eigenvalues of $ H_{\text{eff}} $ with Re$ [E] $ on the $ x $-axis and Im$ [E] $ on the $ y $-axis in Figs.~\ref{fig:Heff}~(c) and (d), and the reflection spectrum (blue line) under the same parameter conditions in the same figure. 
The eigenvalues for $ d=\lambda_0/4 $ are displayed as the red dots in Fig.~\ref{fig:Heff}~(c), where an energy gap emerges around the resonant frequency, remarkably matching the ultrahigh reflection window.
The results for $d$ close to, but not exactly equals to, the Bragg distance (e.g. $ d = 0.024 \lambda_0 $) are presented in Fig.~\ref{fig:Heff}~(d), further demonstrating that ultrahigh reflection emerges within the energy gap region.  
These observations provide compelling evidence that the ultrahigh reflection results from the energy gap engendered by collective interactions among the atoms.

Actually, for infinitely long atom chain, the eigenvalue of  $H_{\text{eff}}$  shown in Eq.~\eqref{eq:effectiveh} in the first excitation subspace can be solved and  it is given by (see Subsec.~\ref{app:subsec_eq-5} in Appendix )
\begin{equation} \label{eq:dispersion_relation}
\omega(K)=\omega_{0}+\frac{\Gamma_0}{2}\frac{\text{sin}(kd)}{\text{cos}(Kd)-\text{cos}(kd)},
\end{equation}
with eigenfunction $|\psi\rangle=N^{-1/2}\sum_{j}e^{iKz_j}\sigma_{j}^{\dagger}|g\rangle$. The relation in Eq.~\eqref{eq:dispersion_relation} is often called as the dispersion relation of the system. For an infinitely long atomic chain, $K$ can vary continuously, and thus $\cos(Kd)\in [-1,1]$. The first excitation subspace has a gap with bandwidth 
\begin{equation} \label{eq:bandgap}
		\Delta \omega_{bg} =\frac{\Gamma_{0}}{2}(\frac{\sin{k d}}{1 - \cos{k d}}-\frac{\sin{k d}}{-1 - \cos{k d}}) =\frac{\Gamma_{0}}{\sin(k d)}.
\end{equation}
Taking $d = 0.25\lambda_0$ as an example, the dispersion relation is shown in Fig.~\ref{fig:Heff}~(e) which clearly demonstrates the existence of a band gap with width $ \Gamma_{\mathrm{0}}$. This band gap  exactly matches to the bandwidth of the ultrahigh-reflectivity region. In Fig.~\ref{fig:Heff}~(f) we present the band gap (pink area) for different values of atom separation $d$. We can see that the band gap diverges when $d$ is close to the Bragg distance and it matches the ultrahigh reflection window shown in Fig.~\ref{fig:Heff}~(a) (see black dashed curve)
When $d=\lambda_0/4$, the band gap $\Delta\omega_{bg}=\Gamma_{0}$, which could be increased to $6.66\Gamma_{0}$ when $d=0.024\lambda_0$.

\section{Extend the bandwidth of ultrahigh reflection window}

\begin{figure}[htbp]
    \centering
    \includegraphics[width=1.0\linewidth]{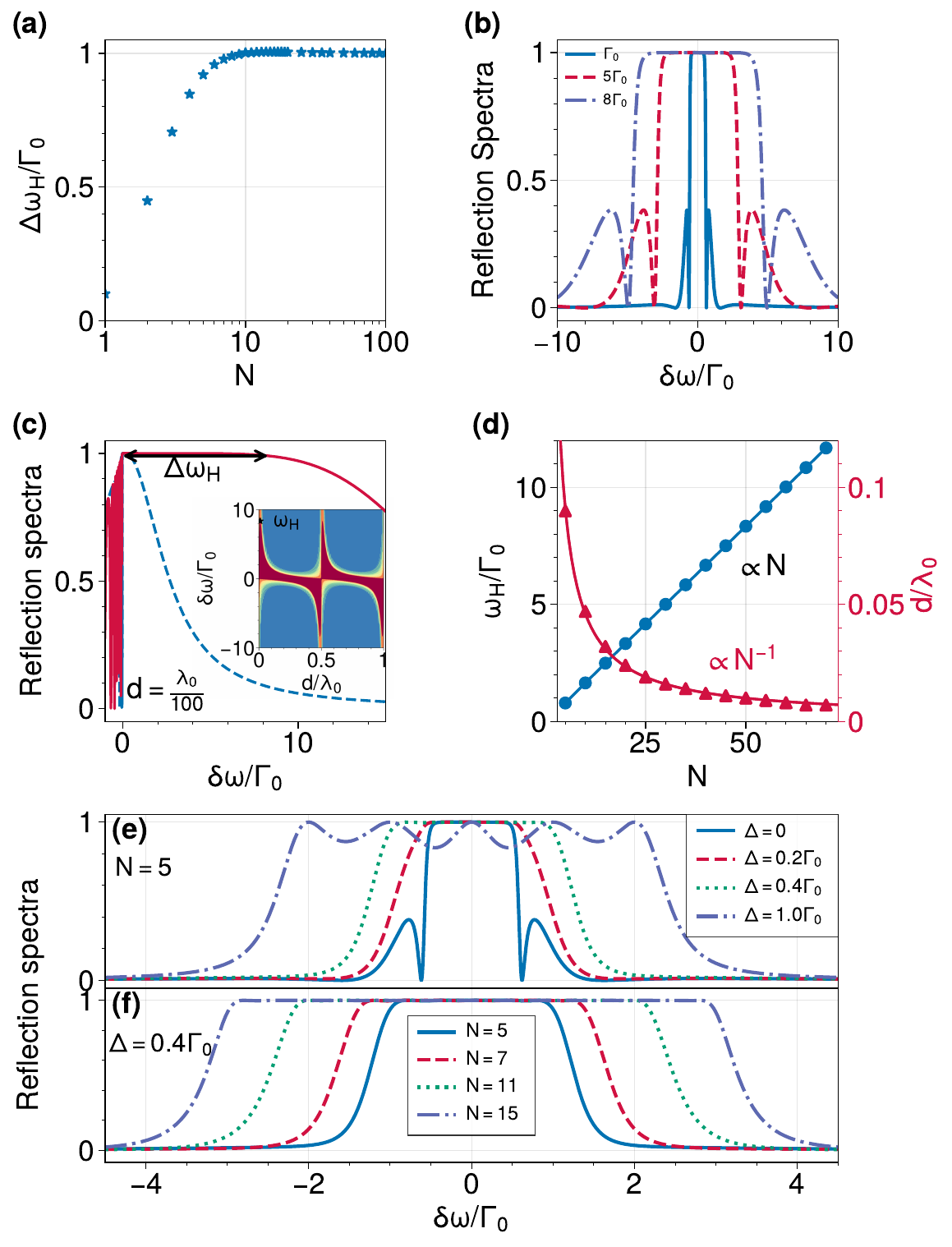}
    \caption{  (a) The ultrahigh reflection bandwidth ($ R \geq 99 \% $) as a function of atom number when $ d =\lambda_{0}/4 $. (b) The reflection spectra for three different values of $ \Gamma_{\text{1D}} $ when the atom number is fixed to be $ N = 5 $.      (c) Reflection spectra when the atom separation is close to the Bragg distance ($ \lambda_0/100 $) when $N=5$ (blue dashed lines) and $ N=50 $ (red solid line). The inset shows the reflection spectra with varying atom separation when $N=50$.  
    (d) The largest total reflection bandwidth as a function of atom number $N$ and the corresponding atom separation. (e) Reflection spectra for $ N = 5 $ with different frequency modulations. (f) Reflection spectra of atomic chains with fixed frequency modulation (i.e., $ \Delta = 0.4 \Gamma_0 $) for different number of atoms.
    }
    \label{fig:na_50_k}
\end{figure}  

In Sec.~\ref{subsec:anti_Bragg} we see that there is a finite ultrahigh reflection window around the resonant frequency when $d=\lambda_0/4$ and this window can increase when the number of atom increases. In Fig. ~\ref{fig:na_50_k}~(a), we plot the width of the ultrahigh reflection window $\Delta\omega_{H}$ as a function of $N$ when $d=\lambda_0/4$. The result shows that although the width $\Delta\omega_{H}$ indeed increases with $N$ when $N$ is small, it quickly saturates when $N$ is large and the ultimate width is limited to $\Gamma_{0}$. Thus, to further increase the ultrahigh reflection window when $d=\lambda_0/4$, we have to increase the value of $\Gamma_{\mathrm{1D}}$ as shown in Fig. ~\ref{fig:na_50_k}~(b) where the reflection spectra for three different values of $\Gamma_{\mathrm{1D}}$ are presented. However, it is not easy to increase $\Gamma_{\mathrm{1D}}$ in practice. We thus seek for alternative methods to extend the ultrahigh reflection window.  


As shown in Fig.~\ref{fig:Heff}~(a), the ultrahigh reflection window is significantly large when $d$ is close to but not equal to the Bragg distance. When the number of atoms increases, e.g. $N=50$, the reflection spectrum is shown in the inset of Fig. ~\ref{fig:na_50_k}~(c) from which we can see that the ultrahigh reflection window significantly increases near the Bragg distance. 
%
For example, under the specific atomic separation condition of $ d = \lambda_0 /100 $, we compare the reflection spectra for two distinct atomic configurations: $ N = 5 $ (blue dashed line) and $ N = 50 $ (red solid line) in Fig.~\ref{fig:na_50_k}~(c).
It is evident that the ultrahigh reflection window is substantially enlarged as $ N $ increases from $ 5 $ to $ 50 $. 
Specifically, for $ N = 5 $, the ultrahigh reflection bandwidth $ \Delta \omega_{\mathrm{H}} $ is approximately $ 0.79 \Gamma_{0}$. In contrast, for $ N = 50 $, $ \Delta \omega_{\mathrm{H}} \approx 8.34 \Gamma_{0}$, indicating an almost tenfold increase in the ultrahigh reflection bandwidth.  
Additionally, Fig.~\ref{fig:na_50_k}~(d) illustrates the maximum bandwidth of the ultrahigh reflection window ($ R > 0.99 $) for varying numbers of atoms, revealing a nearly linear relationship between the bandwidth and $ N $ (blue dots). 
Moreover, the atomic separation leading to the maximum bandwidth also varies with $N$ (red triangles), while the the largest ultrahigh reflection bandwidth corresponds to the separation closest to the Bragg distance. 
For instance, when $ N = 60 $, the bandwidth of the ultrahigh reflection window reaches approximately $ 10 \Gamma_{0} $ at $ d = \lambda_0/125 $. 

Another effective strategy to broaden the ultrahigh reflection window is by extending the range of the atomic transition frequencies. This can be achieved by applying an external electromagnetic field to induce a gradient modulation of transition frequencies among the atoms via the Stark or Zeeman effects, as delineated in Fig.~\ref{fig:model}. Without loss of generality, we assume that $ N $ is an odd number, and the transition frequency of the $ j $th atom is given by $ \omega_{a}+(\frac{N+1}{2}-j)\Delta$, where $\omega_{a} $ is the transition frequency of the central atom, and $ \Delta $ represents the energy difference between two adjacent atoms.  
By introducing a transition energy difference of $\Delta$ between two adjacent atoms and adopting $ N = 5 $ and $ d=\lambda_0/4 $ as a paradigm, we compare the reflection spectra for different values of $ \Delta $, as shown in Fig.~\ref{fig:na_50_k}~(e). Indeed, the ultrahigh reflection window evidently expands as $ \Delta $ increases (e.g., $ \Delta = 0.2\Gamma_{0} $ (red dashed line) with $ \Delta = 0.4 \Gamma_{0} $ (green dotted line)).
This broadened bandwidth of ultrahigh reflection is a consequence of the constructive interference among reflections from different atoms. 
However, $ \Delta $ cannot be arbitrarily large. For example, when $ \Delta = \Gamma_{0} $, the reflection curve within the ultrahigh reflection window no longer maintains a flat profile, exhibiting several dips where reflectivity experiences significant declines. 
This phenomenon arises because, when the energy difference between atoms approaches or exceeds the emission linewidth, the interference effects are substantially diminished. 
Therefore, it is critical to select an appropriate energy shift. 
By setting $ \Delta = 0.4 \Gamma_0 $, we further examine the reflection spectra for varying $N$, with results depicted in Fig.~\ref{fig:na_50_k}~(f). 
It is clearly observable that the ultrahigh reflection window can be substantially enlarged as $N$ increases. 
Thus, this method allows us to finely tune the bandwidth of the ultrahigh reflection window, which could be utilized as an optical switch.

\section{\label{sec:dissipation}Effects of external dissipation}

\begin{figure}[htbp]
    \centering
    \includegraphics[width=1.0\linewidth]{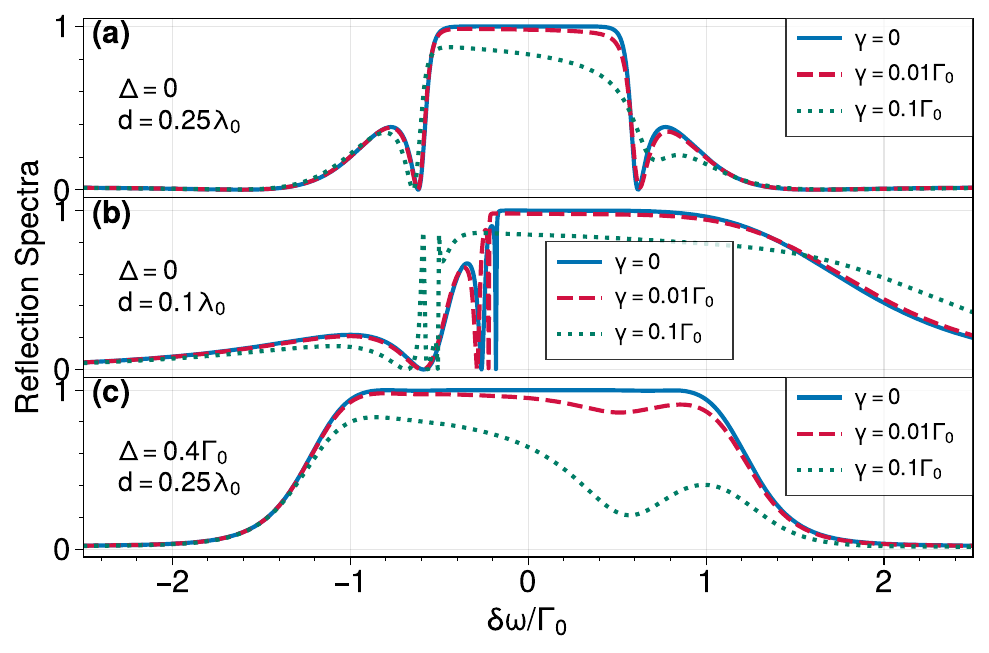}
    \caption{Reflection spectra for a $ 5 $-atom waveguide dissipate outside of the waveguide with varying decay rates $ \gamma $ spanning from ideal to non-ideal scenarios: (a) $ d = 0.25 \lambda_0 $, no frequency modulation; (b) $ d = 0.1 \lambda_0 $, no frequency modulation; (c) $ d = 0.25 \lambda_0 $, with frequency modulation $ \Delta = 0.4 \Gamma_0 $. 
    }
    \label{fig:dissipation}
\end{figure}  
In the preceding discussions, we have neglected the external dissipation to simplify the model.
Here, we investigate non-ideal scenarios where atoms can dissipate part of their energy into non-waveguide modes (i.e, $ \gamma_j \neq 0 $). 
Taking $N=5$ atoms as an example, we examine three distinct cases: symmetric reflection spectra at $ d = \lambda_0/ 4 $ ( Fig.~\ref{fig:dissipation}~(a)), asymmetric reflection spectra at $ d = \lambda_0/10 $ ( Fig.~\ref{fig:dissipation}~(b)), and extended reflection spectra under frequency modulation (Fig.~\ref{fig:dissipation}~(c)). 
We compare the reflection spectra across three different external dissipation rates: $ \gamma = 0 $, $ 0.01 \Gamma_0 $, and $ 0.1 \Gamma_0 $. 
As anticipated, the presence of external dissipation attenuates reflectivity, especially when $ \gamma $ is large. 
However, for $ \gamma = 0.01 \Gamma_0 $, which is experimentally attainable~\cite{Scarpelli2019,mirhosseini2019}, the reflectivity within the ultrahigh reflection window can still be larger than $ 91.5\% $ across all three scenarios.
Therefore, the ultrahigh reflection with broad bandwidth should be observable under current technology.

\section{\label{sec:conclusion}Conclusion}
In conclusion, we present a comprehensive investigation of the broadband reflection properties exhibited by a single photon propagating through an array of atoms coupled to a 1D waveguide.
The mechanism underlying these flat bands with ultrahigh reflectivities can be interpreted as a consequence of the energy band gap formed by the collective atom-atom interactions.
We propose several strategies to further broaden the ultrahigh reflection windows, 
including increasing the number of atoms with separations near the Bragg distance,
and introducing a gradient external electric field to induce gradient frequency modulation among the atoms.  The center frequency and bandwidth of the broadened ultrahigh reflection window is dynamically adjustable here.
Additionally, we examine the effects of external dissipation, and demonstrate that ultrahigh reflection with a broad bandwidth should remain observable in waveguide-QED systems with relatively modest levels of external dissipation. 
These high-bandwidth designs pave the way for innovative avenues in broadband local photonic processing and hold immense potential for the implementation of chip-integrated band filter, quantum storage, optical switching, and wavelength-selective devices.

\begin{acknowledgments}
    This work was supported by the National Key R$\&$D Program of China (Grant No. 2021YFA1400800), the Key Program of National Natural Science Foundation of China (Grant No. 12334017), the Key Area Research and Development Program of Guangdong Province (Grant No.2018B030329001), the Guangdong Basic and Applied Basic Research Foundation (Grant No.2023B1515040023), and the Natural Science Foundations of Guangdong (Grant No.2021A1515010039).
\end{acknowledgments}

\appendix

\section{\label{app:subsec-A}Derivation of the reflection spectrum}
\subsection{Derivation of Eq.~(2) in the main text}

Consider that a single photon with wave vector $ k $ is injected from the left end of the waveguide, while all the atoms are initially in the ground state. In this case, the quantum state at arbitrary time can be expressed as
\begin{equation}     \label{eq:quantum-state}
    \ket{\psi(t)}=\sum_{j=1}^{N}\alpha_j(t) \mathrm{e}^{-i \omega_j t} \ket{e_j,0} +\sum_k\beta_k(t)\mathrm{e}^{-i \omega_k t}\ket{g,1_k}
    \,,
\end{equation}
where $ \ket{e_j,0} $ denotes that only the $ j $th atom is in the excited state with null photon in the waveguide and $ \alpha_{j}(t) $ represents its corresponding amplitude. The state $ \ket{g,1_k} $ indicates the presence of a single photon with wave vector $ k $ in the waveguide, while all atoms remain in their ground state, characterized by the amplitude $ \beta_{k}(t) $.

By employing the Schr\"{o}dinger equation $i\hbar|\dot{\psi}(t)\rangle=H|\psi(t)\rangle$ with Hamiltonian given by Eq. (1) in the main text, we can derive the dynamics of the emitter excitation $ \alpha_{j}(t) $, as well as the photon amplitude $ \beta_{k}(t) $~\cite{liao2016c}:
\begin{equation} \label{eq:alpha}
    \dot{\alpha}_j(t)=-i \sum_k g_k^{j} e^{i k z_j} \beta_k(t) e^{-i \delta \omega_k^{j} t}-\frac{\gamma_{j}}{2} \alpha_j(t) 
\end{equation}
\begin{equation} \label{eq:beta}
    \dot{\beta}_k(t)=-i \sum_{j=1}^{N} g_k^{j*} e^{-i k z_j} \alpha_j(t) e^{i \delta \omega_k^{j} t}
\end{equation}
where $\delta \omega_k^{j} \equiv \omega_k-\omega_j=\left(|k|-k_j\right) v_g$ is the detuning between the atomic transition frequency and the frequency of the guided photon.

Integrating Eq.~\eqref{eq:beta} we obtain the photon amplitude formula
\begin{equation}  \label{eq:beta1}
    \beta_k(t)=\beta_k(0)-i \sum_{j=1}^{N} g_k^{j*} e^{-i k z_j} \int_0^t \alpha_j\left(t^{\prime}\right) e^{i \delta \omega_k^{j} t^{\prime}} d t^{\prime}
\end{equation}
where $\beta_k(0)$ is the initial photon amplitude. On substituting from  Eq.~\eqref{eq:beta1} into  Eq.~\eqref{eq:alpha}, the dynamics of the atomic excitations is
\begin{equation}  \label{eq:alpha1}
    \begin{aligned}
        \dot{\alpha}_j(t)&=-  i \sum_k g_k^{j} e^{i k z_j} \beta_k(0) e^{-i \delta \omega_k^{j} t}-\frac{\gamma_{j}}{2} \alpha_j \\
         &-\sum_{l=1}^{N} \sum_k g_k^{j} g_{k}^{l*} e^{-i k\left(r_j-r_l\right)} \int_0^t \alpha_l\left(t^{\prime}\right) e^{i \delta \omega_k^{j} t^{\prime}} d t^{\prime} e^{-i \delta \omega_k^{j} t}
         \,.
        \end{aligned}
\end{equation}

Replace the summation over $k$ by integration for a long 1D waveguide
\begin{equation}
    \sum_k \rightarrow \frac{L}{2 \pi} \int_{-\infty}^{\infty} d k
\end{equation}
where $L$ is the quantization length in the propagation direction.
After Weisskopf-Wigner approximation, Eq.~\eqref{eq:alpha1} turns into
\begin{equation}  \label{eq:atom-evolution}
    \begin{aligned}
        \dot{\alpha}_j(t)&=b_j(t) \\
        &-\sum_{l=1}^{N}\left(\frac{\sqrt{\Gamma_{j}\Gamma_{l}}}{2} \mathrm{e}^{i k_l z_{j l}}+\frac{\gamma_{j}}{2} \delta_{j l}\right) \alpha_l\left(t-\frac{z_{j l}}{v_g}\right)e^{i\Delta_{jl}t}
        \,,   
    \end{aligned}
\end{equation}
with 
\begin{equation} 
b_j(t)=-\frac{i}{2 \pi} \sqrt{\frac{\Gamma_{j} v_g L}{2}} \int_{-\infty}^{\infty} \beta_k(0) \mathrm{e}^{i k z_j-i \delta \omega_k^{j} t} \mathrm{d} k
\,.
\end{equation}
Here, $ b_j(t) $ serves as the pump source of photon on atoms. It is important to note that the excitation probability of the $ l $th atom influences that of the $ j $th atom, reflecting the dipole-dipole interactions between atoms situated at different locations. It is clearly seen that the effective interaction between atoms in a single-mode waveguide is a long-range interaction. 
The decay rate $\Gamma_{j}=2 L\left|g_{k_a}^{j}\right|^2 / v_g$ couples the atoms inside the waveguide, and $ z_{j l} = z_j - z_l $ is the distance between the $ j $th and $ l $th atom. $\Delta_{jl}=\omega_{j}-\omega_{l}$.
In Eq.~\eqref{eq:atom-evolution}, the term $ \Gamma_{j} / 2 $ does not attenuate with distance, relating to the intensity of the interaction between different atoms.

Supposing that the incident photon pulse is propagating in the positive $k$ direction,  we can obtain the amplitude of the reflection and transmission photon modes in the long-time limit from  Eq.~\eqref{eq:beta1} which are given by
\begin{equation}
    \begin{aligned}
    &\beta_R(\delta k)\\
    &=-i  \sum_{j=1}^{N} \sqrt{\frac{\Gamma_{j} v_g}{2 L}}e^{i(k_0+\delta k) z_j} \int_0^\infty \alpha_j\left(t^{\prime}\right) e^{i (\delta k-\Delta k_j) v_g t^{\prime}} d t^{\prime}, 
    \end{aligned}
\end{equation}
\begin{equation}
    \begin{aligned}
        &\beta_T(\delta k)=\beta_{\delta k}(0) \\
    &-i  \sum_{j=1}^{N} \sqrt{\frac{\Gamma_{j} v_g}{2 L}}e^{-i(k_0+\delta k) z_j} \int_0^\infty \alpha_j\left(t^{\prime}\right) e^{i (\delta k-\Delta k_j) v_g t^{\prime}} d t^{\prime}, 
    \end{aligned}
\end{equation}
where $\delta k=|k|-k_0$ and $\Delta k_j=k_j-k_0$ with $k_0$ being a reference positive wavevector.



%

%
Define the integration term in Eq.~\eqref{eq:beta}
\begin{equation} \label{eq:chi}
    \chi_j(\delta k)=\int_{-\infty}^{\infty} \alpha_j(t) e^{i \delta k v_g t} d t
\end{equation}
where $\alpha_j(t)=0$ for $t<0$ with $j=1,2, \ldots, N$. By Fourier transformation, Equation \eqref{eq:chi} changes into
\begin{equation} \label{eq:alpha2}
    \alpha_j(t)=\frac{v_g}{2 \pi} \int_{-\infty}^{\infty} \chi_j(\delta k) e^{-i \delta k v_g t} d \delta k .
\end{equation}
After inserting Eq.~\eqref{eq:alpha2} into Eq.~\eqref{eq:atom-evolution}, we finally have the amplitude of the reflected and the  transmitted photon modes:
\begin{equation} \label{eq:refl}
        \beta_{R}(\delta k)=-i\sum_{j=1}^{N}\sqrt{\frac{\Gamma_{j} v_g}{2L}}\mathrm{e}^{i (k_0+\delta k) z_j}\chi_j(\delta k-\Delta k_j),
\end{equation}
\begin{equation} \label{eq:trans}
        \beta_T(\delta k) =\beta_{\delta k}(0)-i\sum_{j=1}^{N}\sqrt{\frac{\Gamma_{j} v_g}{2L}}\mathrm{e}^{-i(k_0+\delta k)z_j}\chi_j(\delta k-\Delta k_j). 
\end{equation}
By performing Fourier transformation on both side of Eq. (S7), we can obtain \cite{liao2016c}
\begin{equation}
    \chi_j(\delta k-\Delta k_j)=-i\sum_{l=1}^{N}\sqrt{\frac{\Gamma_{l}L}{2v_{g}}} M_{j l}^{-1}(\delta k) \beta_l(\delta k)e^{i(k_0+\delta k)z_{j}}
\end{equation}
with 
\begin{equation}
    M_{jl}(\delta k)=\frac{\sqrt{\Gamma_j \Gamma_l}}{2}\mathrm{e}^{ik z_{jl}}+[\frac{\gamma_{j}} {2}-i (\delta k-\Delta k_{j}) v_g]\delta_{jl} 
    \,.
\end{equation}
Supposing that the incident photon is a plane wave with $ \beta_{\delta k}(0) = 1 $, the reflection amplitude can be simplified as
\begin{equation} \label{eq:r_omega0}
	r(\omega) =\beta_{R}(\omega)= -\frac{1}{2} \sum_{j = 1}^{N} \sum_{l = 1}^{N} \sqrt{\Gamma_j \Gamma_l} {\rm{e}}^{i k (z_j + z_l )} M_{jl}^{-1}(\omega)
\end{equation}
with $\omega=kv_{g}$ and $M_{jl}(\omega) = (\sqrt{\Gamma_j \Gamma_l}/2) {\rm{e}}^{i k |z_j - z_l |} +[\gamma_j / 2 - i (\omega - \omega_j)] \delta_{jl}$.



\subsection{\label{app:subsec_eq-4}Derivation of Eq.~(4) in the main text}
Now we derive Eq.~(4) in the main text, with the consideration of identical atoms (i.e., $\omega_{j}=\omega_{0}, \Gamma_{j}=\Gamma_{0}$, and $\gamma_{j}=\gamma$). 
The refection amplitude shown in Eq.~\eqref{eq:r_omega0} can be expressed in the form of the matrix components $ M_{jl}^{-1}(\omega) $ multiplied by the vector components $\psi_{in,j(l)}$: 
\begin{equation} \label{eq:r_omega1}
	r(\omega) = - \frac{\Gamma_{\text{1D}}}{2}\sum_{j = 1}^{N} \sum_{l = 1}^{N} \psi_{in,j} M_{jl}^{-1}(\omega) \psi_{in,l}
\end{equation}
while $\psi_{in,j(l)}$ is the $ j(l) $th vector component with $\vec{\psi}^T_{in} = (\mathrm{e}^{i k z_1}, \mathrm{e}^{i k z_2}, \cdots ,\mathrm{e}^{i k z_N})$, representing the phase of the incident field at each atomic position.

Rewrite Eq.~\eqref{eq:r_omega1} into the multiplication of the defined matrices:
\begin{equation} \label{eq:r_omega2}
	r(\omega) = - i \frac{\Gamma_{\text{1D}}}{2}\vec{\psi}^T_{in} (i\mathbf{M})^{-1} \vec{\psi}_{in}
\end{equation}
with
\begin{equation} \label{eq:iM1}
	i\mathbf{M} =  \omega \mathbf{I} - [(\omega_0 - i \gamma / 2 )\mathbf{I} + \mathbf{g}]
    \,,
\end{equation}
while $ \mathbf{I} $ is the unit matrix. The element of the matrix $\mathbf{g}$ is given by $g_{ij} = (-i\Gamma_{\text{1D}} /2) {\rm{e}}^{i k |z_j - z_l|}$. 
We have already multiplied $ i $ to the numerator and denominator in Eq.~\eqref{eq:r_omega2} to conform to the form of effective Hamiltonian in the main text (Eq.~(3)). Recall $ H_{\text{eff}} $ here:

In the single-excitation subspace, substitute Eq.~\eqref{eq:effective_H} into Eq.~\eqref{eq:iM1} we have
\begin{equation} \label{eq:iM2}
	i\mathbf{M} = \omega \mathbf{I} - H_{\text{eff}}
\end{equation}
On account of $H_{\text{eff}}$ is a complex symmetric matrix, it can be diagonalized into $H_{\text{eff}} \vec{v}_\xi = \lambda_\xi \vec{v}_\xi  \; (\xi \in 1, \cdots , N)$, with eigenvectors $ \vec{v}_\xi $ and eigenvalues $ \lambda_\xi $. The eigenvectors satisfy the orthogonal normalization relation:
\begin{equation}
	\sum_{\xi = 1}^N \vec{v}_\xi \otimes \vec{v}^T_\xi = \mathbf{I}, \; \vec{v}^T_\xi \cdot \vec{v}_{\xi^\prime} = \delta_{\xi, \xi^\prime}
    \,.
\end{equation}

Equation.~\eqref{eq:iM2} denotes that as $ \omega \mathbf{I} $ is a constant term, the matrix $i\mathbf{M}$ shares the same set of eigenstates with $H_{\text{eff}}$, thereby $(i\mathbf{M})^{-1}$ can be spectrally decomposed into:
\begin{equation}  \label{eq:iM3}
	(i\mathbf{M})^{-1} = \sum_{\xi = 1}^{N} \frac{1}{\omega-\lambda_\xi} \vec{v}_\xi \otimes \vec{v}^T_\xi
\end{equation}
Substitute Eq.~\eqref{eq:iM3} into Eq.~\eqref{eq:r_omega2}, we have the final reflectivity form (Eq.~(4)):
\begin{equation}
\begin{split}
	r(\omega) & = - i \frac{\Gamma_{\text{1D}}}{2}\sum_{\xi = 1}^{N} \frac{(\vec{\psi}^T_{in} \cdot \vec{v}_\xi) ( \vec{v}^T_\xi \cdot \vec{\psi}_{in} )}{\omega - \lambda_\xi} \\ & = - i \frac{\Gamma_{\text{1D}}}{2}\sum_{\xi = 1}^{N} \frac{(\vec{\psi}^T_{in} \cdot \vec{v}_\xi)^2 }{\omega - \lambda_\xi}
    \end{split}
\end{equation}
with $\vec{\psi}^T_{in} = (\mathrm{e}^{i k z_1}, \mathrm{e}^{i k z_2}, \cdots ,\mathrm{e}^{i k z_N})$.

\subsection{\label{app:subsec_eq-5}Derivation of Eq.~(5) in the main text}
For an infinite atomic chain, assuming the eigenstate to be $\vec{v}_K = (N)^{-1/2}\sum_{n = -\infty}^{\infty} \mathrm{e}^{i K n d} \hat{\sigma}_n^\dagger \left| 0 \right\rangle $.
Act the effective Hamiltonian on it:
\begin{equation}\label{eq:effective_H1}
    \begin{aligned}
        &H_{\mathrm{eff}} \vec{v}_K =   \\
        &\omega_0 \vec{v}_K-i \frac{\Gamma_{\mathrm{1D}}}{2} \frac{1}{\sqrt{N}}\sum_{j,l = -\infty}^{\infty} \hat{\sigma}_j^\dagger \hat{\sigma}_l \mathrm{e}^{i k |j - l| d} \sum_{n = -\infty}^{\infty} \mathrm{e}^{i K n d} \hat{\sigma}_n^\dagger \left| 0 \right\rangle
        \,.
    \end{aligned}
\end{equation}
In this context, $\left| 0 \right\rangle $ represents all atoms in the ground state, where $\left| 0 \right\rangle  \equiv \left| gg\cdots g \right\rangle$. The state $\left| e \right\rangle_l \equiv \hat{\sigma}_l^\dagger\left| 0 \right\rangle$ indicates that the $l$th atom is excited while all other atoms remain in the ground state. For the single-excitation subspace, the following relation will naturally be established:
\begin{equation}
	\hat{\sigma}_l \hat{\sigma}_n^\dagger\left| 0 \right\rangle = \hat{\sigma}_l \left| e \right\rangle_n = \delta_{nl}\left| 0 \right\rangle
    \,.
\end{equation}
Therefore Eq.~\eqref{eq:effective_H1} changes into
\begin{equation}\label{eq:effective_H2}
    \begin{split}	
	    H_{\mathrm{eff}} \vec{v}_K & =\omega_0 \vec{v}_K -i \frac{\Gamma_{\mathrm{1D}}}{2} \frac{1}{\sqrt{N}}\sum_{j,l = -\infty}^{\infty} \mathrm{e}^{i k |j - l| d}  \mathrm{e}^{i K l d} \hat{\sigma}_l^\dagger \left| 0 \right\rangle \\
	    & =\omega_0 \vec{v}_K -i \frac{\Gamma_{\mathrm{1D}}}{2}\frac{1}{\sqrt{N}} \sum_{j = -\infty}^{\infty}\sum_{l = -\infty}^{\infty} \mathrm{e}^{i k |j - l| d} \mathrm{e}^{i K l d} \hat{\sigma}_l^\dagger \left| 0 \right\rangle
    \end{split}
    \,.
\end{equation}
Furthermore, the second term above could be simplified by the following relations 
\begin{equation} 
    \begin{split}
    	& \sum_{l = -\infty}^{\infty} \mathrm{e}^{i k |j - l| d} \mathrm{e}^{i K l d} \\ 
    	& = \sum_{l = -\infty}^{j - 1} \mathrm{e}^{i k (j - l) d} \mathrm{e}^{i K l d} + \sum_{l = j}^{-\infty} \mathrm{e}^{i k (l - j) d} \mathrm{e}^{i K l d} \\
    	& = -\frac{\mathrm{e}^ { i K l d}}{1 - \mathrm{e}^ { i (K - k) d} } + \frac{\mathrm{e}^ { i K l d}}{1 - \mathrm{e}^ { i (K + k) d} } \\
    	& = i \frac{\sin{k d}}{\cos{K d} - \cos{k d}} \mathrm{e}^ { i K l d}
    \end{split}
    \,.
\end{equation}
Substitute it into Eq.~\eqref{eq:effective_H2} we have
\begin{equation} \label{eq:effective_H3}
	\begin{split}	
		&H_{\mathrm{eff}} \vec{v}_K \\
        & =\omega_0 \vec{v}_K -i \frac{\Gamma_{\mathrm{1D}}}{2} \frac{1}{\sqrt{N}}\sum_{j = -\infty}^{\infty} i \frac{\sin{k d}}{\cos{K d} - \cos{k d}} \mathrm{e}^ { i K l d}   \hat{\sigma}_l^\dagger \left| 0 \right\rangle \\
		& =\left(\omega_0 + \frac{\Gamma_{\mathrm{1D}}}{2} \frac{\sin{k d}}{\cos{K d} - \cos{k d}}\right) \vec{v}_K
	\end{split}
\end{equation}
It is noteworthy that, for an infinite chain, all eigenvalues of the effective Hamiltonian are real, as evidenced by Eq.~\eqref{eq:effective_H3}. For a given eigenstate $\vec{v}_K$, the corresponding eigenenergy is expressed as follows:
\begin{equation}
	\omega(K) = \omega_0 + \frac{\Gamma_{\mathrm{1D}}}{2} \frac{\sin{k d}}{\cos{K d} - \cos{k d}}
\end{equation}

\section{\label{app:subsec-B} Reflection spectrum by multiple-scattering theory and the transfer matrix theory}
For a single atom with transition frequency $ \omega_j $, the reflection and transmission amplitudes are given by \cite{liao2015}
\begin{equation}
    r_{j}(k) =-\frac{\Gamma_{\text{1D}}}{\Gamma_{\text{1D}}+\gamma_{j}-2i(k-k_j)v_g}
    \label{eq:rhoj}
\end{equation}
\begin{equation}
    t_{j}(k)=-\frac{\gamma_{j}-2i(k-k_j)v_g}{\Gamma_{\text{1D}}+\gamma_{j}-2i(k-k_j)v_g}
    \,,
    \label{eq:tranj}
\end{equation}
respectively. It is evident that at resonance, without external dissipation ($ \gamma_{j} = 0 $), the photon is completely reflected with a $ \pi $ phase shift. 

Just like a Fabry–Pérot interferometer (FPI), which is an optical cavity formed by two parallel reflecting surfaces (i.e.: thin mirrors), two atoms can also act as mirrors, thereby constituting an atomic FPI. Assuming an incident electromatic field $ E_{\text{in}} $ approaches from the left, the reflectivity of this interaction can be computed as follows:
    
\begin{equation}
    \begin{aligned}
        &r = \frac{E_{\text{refl}}}{E_{\text{in}}}  = r_1  + t_1 \mathrm{e}^{i k d} r_2 \mathrm{e}^{i k d} t_1  +  t_1 \mathrm{e}^{i k d} r_2 \mathrm{e}^{i k d} r_1 \mathrm{e}^{i k d} r_2 \mathrm{e}^{i k d} t_1 \\
        & +  t_1 \mathrm{e}^{i k d} r_2 \mathrm{e}^{i k d} r_1 \mathrm{e}^{i k d} r_2 \mathrm{e}^{i k d} r_1 \mathrm{e}^{i k d} r_2 \mathrm{e}^{i k d} t_1  \\
        & +  t_1 \mathrm{e}^{i k d} r_2 \mathrm{e}^{i k d} (r_1 \mathrm{e}^{i k d} r_2 \mathrm{e}^{i k d})^N t_1 \\
        & = r_1+ \frac{t_1^2 r_2 \mathrm{e}^{2i k d}}{1-r_1 r_2  \mathrm{e}^{2i k d}} \\
    \end{aligned}
\end{equation}

Notably, it is assumed that the reflectivity and transmittivity for an incident pulse from any direction towards the atom possess identical values $ r $ and $ t $ , which indicates that the reflectivity from the left $ r_+ $ and it from the right $ r_- $ are equal $ (r_+ = r_-) $, so as to the transmittivity. Consider the two-atom FPI as a compact system, the reflectivity is 
\begin{equation}
    r = r_1+ \frac{t_1^2 r_2 \mathrm{e}^{2i k d}}{1-r_1 r_2  \mathrm{e}^{2i k d}}
    \label{eq:R-twoatom}
\end{equation}

If the atoms are arranged in a periodic array, the system, which is in the linear response regime for a single photon, resembles a distributed Bragg reflector (DBR) characterized by the stacking of multiple layers of alternating materials with periodically high and low effective refractive indices~\cite{y.d.chong2010}.  Thereby,  we could derive the reflection amplitude of a $ N $-atom system using the following recurrence relation of the two-atom system~\cite{sophoclesjorfanidis2002}.
\begin{equation}
    R_j(k)=r_j(k)+ \frac{t_j(k)^2 R_{j+1}(k) \mathrm{e}^{2i k_j d}}{1-r_j(k) R_{j+1}(k) \mathrm{e}^{2i k_j d}}
    \,, 
    \label{eq:R-EWT}
\end{equation}
where $ R_j(k) $ ($ j=N-1, \ldots , 2, 1 $) represents the aggregate reflectivity for an incident single photon at the $ j $th atom (reflector) situated at position $ z_j $, with the boundary condition $ R_{N}(k)=r_{N}(k) $. The reflectivity and transmittance $ r_{j}(k) $ and $ t_j(k) $ are given by Eqs.~\eqref{eq:rhoj} and ~\eqref{eq:tranj}, respectively. By iteratively calculating the reflection amplitude $ R_{j}(k) $ starting from $ R_{N}(k) $, we ultimately derive $ R_{1}(k) $, which represents the total reflection amplitude of the entire atomic array.
From Eq.~\eqref{eq:refl} and Eq.~\eqref{eq:R-EWT}, it is worth noting that the total reflectivity is significantly influenced by the atom separation $ d $. 

Another straightforward way to derive the reflectivity and transmittivity of light propagation through a one-dimensional atomic chain, is the transfer matrix method~\cite{corzo2016}.
For an atom at position $z$, the matching matrix is written to relate the fields $ E_{\pm} $ on the left of the atom to the fields $ E_{\pm}^{\prime} $ on the right:
\begin{equation}
    \left[
        \begin{array}{l}
            E_{+} \\
            E_{-}
            \end{array}\right]
            =\frac{1}{t}
                \left[\begin{array}{cc}
                t^2-r^2 & r \\
                -r & 1
                \end{array}\right]
            \left[\begin{array}{l}
            E_{+}^{\prime} \\
            E_{-}^{\prime}
        \end{array}
    \right] \quad \text { (matching matrix) }
    \,.
\end{equation}
Express the matching matrix with $ M_a $:
\begin{equation}
    \mathbf{M}_a=\frac{1}{t}\left[\begin{array}{cc}
        t^2-r^2 & r \\
        -r & 1
        \end{array}\right]
\end{equation}
 where $t$ and $r$ are single-atom transmission and reflection coefficients respectively. The propagation matrix between two neighboring atoms separated by a distance $ d $ is given by
\begin{equation}
    \left[\begin{array}{l}
    E_{1+} \\
    E_{1-}
    \end{array}\right]
    =\left[\begin{array}{cc}
    \mathrm{e}^{j k d} & 0 \\
    0 & \mathrm{e}^{-j k d}
    \end{array}\right]\left[\begin{array}{l}
    E_{2+} \\
    E_{2-}
    \end{array}\right] \quad \text { (propagation matrix) }
\end{equation}
where $ E_{1\pm} $ is the fields on the left of the first atom, and $ E_{2\pm} $ is the fields on the left of the second one. 
Express the propagation matrix with $ M_p $:
\begin{equation}
    \mathbf{M}_p=\left[\begin{array}{cc}
        \mathrm{e}^{i k d} & 0 \\
        0 & \mathrm{e}^{-i k d}
        \end{array}\right]
        \,.
\end{equation}

 The transfer matrix for the full ensemble is obtained as a product of matrices $M_a$ and $M_p$ as follows:
 \begin{equation}
    \mathbf{M}=\left[\begin{array}{ll}
        \mathbf{M}_{11} & \mathbf{M}_{12} \\
        \mathbf{M}_{21} & \mathbf{M}_{22}
        \end{array}\right]=\left(\mathbf{M}_a \cdot \mathbf{M}_p\right)^N
 \end{equation}

 Thereby, the whole interaction process obeys the following transfer matrix: 
\begin{equation}
    \left[
        \begin{array}{l}
            E_{1+} \\
            E_{1-}
            \end{array}\right]
            =\mathbf{M}
            \left[\begin{array}{l}
            E_{N+}^{\prime} \\
            E_{N-}^{\prime}
        \end{array}
    \right] \quad \text { (transfer matrix) }
\end{equation}

The transmission and reflection coefficients of the atomic chain are given by:
\begin{equation}
    \begin{aligned}
        T & =\left|\frac{1}{\mathbf{M}_{22}}\right|^2 \\
        R & =\left|\frac{\mathbf{M}_{12}}{\mathbf{M}_{22}}\right|^2
        \end{aligned}
\end{equation}

For the atomic FPI, we derive the reflectivity and transmittivity employing the transfer matrix formalism:
\begin{equation}
    \mathbf{M}=\frac{1}{t_1}\left[\begin{array}{cc}
        t_1^2-r_1^2 & r_1 \\
        -r_1 & 1
        \end{array}\right]
        \left[\begin{array}{cc}
            \mathrm{e}^{i k d} & 0 \\
            0 & \mathrm{e}^{-i k d}
            \end{array}\right]
        \frac{1}{t_2}\left[\begin{array}{cc}
            t_2^2-r_2^2 & r_2 \\
            -r_2 & 1
            \end{array}\right]
\end{equation}
then
\begin{equation}
    M_{11} = \frac{t_1^2-r_1^2}{t_1} \mathrm{e}^{i k d}   \frac{t_2^2-r_2^2}{t_2} +  \frac{r_1}{t_1} \mathrm{e}^{-i k d}  \frac{-r_2}{t_2} 
\end{equation}
\begin{equation}
    M_{12} = \frac{(t_1^2-r_1^2) r_2}{t_1 t_2} \mathrm{e}^{i k d} + \frac{r_1}{t_1 t_2} \mathrm{e}^{-i k d}  
\end{equation}
\begin{equation}
    M_{21} = \frac{-r_1}{t_1} \mathrm{e}^{i k d}  \frac{t_2^2-r_2^2}{t_2}  + \frac{1}{t_1} \mathrm{e}^{-i k d} \frac{-r_2}{t_2} 
\end{equation}
\begin{equation}
    M_{22} = \frac{-r_1}{t_1} \mathrm{e}^{i k d}  \frac{r_2}{t_2} + \frac{1}{t_1} \mathrm{e}^{-i k d} \frac{1}{t_2} = - \frac{r_1 r_2}{t_1 t_2} \mathrm{e}^{i k d}  + \frac{1}{t_1 t_2} \mathrm{e}^{-i k d} 
\end{equation} 
Therefore, the reflection coefficient is
\begin{equation}
        r = \frac{M_{12}}{M_{22}}=  \frac{ r_1 +  (t_1^2-r_1^2) r_2 \mathrm{e}^{ 2 i k d} }{ 1 - r_1 r_2 \mathrm{e}^{ 2 i k d} } \\
\end{equation}
which is the same with Eq.~\eqref{eq:R-twoatom}.


\begin{thebibliography}{72}%
	\makeatletter
	\providecommand \@ifxundefined [1]{%
		\@ifx{#1\undefined}
	}%
	\providecommand \@ifnum [1]{%
		\ifnum #1\expandafter \@firstoftwo
		\else \expandafter \@secondoftwo
		\fi
	}%
	\providecommand \@ifx [1]{%
		\ifx #1\expandafter \@firstoftwo
		\else \expandafter \@secondoftwo
		\fi
	}%
	\providecommand \natexlab [1]{#1}%
	\providecommand \enquote  [1]{``#1''}%
	\providecommand \bibnamefont  [1]{#1}%
	\providecommand \bibfnamefont [1]{#1}%
	\providecommand \citenamefont [1]{#1}%
	\providecommand \href@noop [0]{\@secondoftwo}%
	\providecommand \href [0]{\begingroup \@sanitize@url \@href}%
	\providecommand \@href[1]{\@@startlink{#1}\@@href}%
	\providecommand \@@href[1]{\endgroup#1\@@endlink}%
	\providecommand \@sanitize@url [0]{\catcode `\\12\catcode `\$12\catcode `\&12\catcode `\#12\catcode `\^12\catcode `\_12\catcode `\%12\relax}%
	\providecommand \@@startlink[1]{}%
	\providecommand \@@endlink[0]{}%
	\providecommand \url  [0]{\begingroup\@sanitize@url \@url }%
	\providecommand \@url [1]{\endgroup\@href {#1}{\urlprefix }}%
	\providecommand \urlprefix  [0]{URL }%
	\providecommand \Eprint [0]{\href }%
	\providecommand \doibase [0]{https://doi.org/}%
	\providecommand \selectlanguage [0]{\@gobble}%
	\providecommand \bibinfo  [0]{\@secondoftwo}%
	\providecommand \bibfield  [0]{\@secondoftwo}%
	\providecommand \translation [1]{[#1]}%
	\providecommand \BibitemOpen [0]{}%
	\providecommand \bibitemStop [0]{}%
	\providecommand \bibitemNoStop [0]{.\EOS\space}%
	\providecommand \EOS [0]{\spacefactor3000\relax}%
	\providecommand \BibitemShut  [1]{\csname bibitem#1\endcsname}%
	\let\auto@bib@innerbib\@empty
	\bibitem [{\citenamefont {Raimond}\ \emph {et~al.}(2001)\citenamefont {Raimond}, \citenamefont {Brune},\ and\ \citenamefont {Haroche}}]{raimond2001}%
	\BibitemOpen
	\bibfield  {author} {\bibinfo {author} {\bibfnamefont {J.~M.}\ \bibnamefont {Raimond}}, \bibinfo {author} {\bibfnamefont {M.}~\bibnamefont {Brune}},\ and\ \bibinfo {author} {\bibfnamefont {S.}~\bibnamefont {Haroche}},\ }\bibfield  {title} {\bibinfo {title} {Manipulating quantum entanglement with atoms and photons in a cavity},\ }\href {https://doi.org/10.1103/RevModPhys.73.565} {\bibfield  {journal} {\bibinfo  {journal} {Rev. Mod. Phys.}\ }\textbf {\bibinfo {volume} {73}},\ \bibinfo {pages} {565} (\bibinfo {year} {2001})}\BibitemShut {NoStop}%
	\bibitem [{\citenamefont {Kimble}(2008)}]{kimble2008}%
	\BibitemOpen
	\bibfield  {author} {\bibinfo {author} {\bibfnamefont {H.~J.}\ \bibnamefont {Kimble}},\ }\bibfield  {title} {\bibinfo {title} {The quantum internet},\ }\href {https://doi.org/10.1038/nature07127} {\bibfield  {journal} {\bibinfo  {journal} {Nature}\ }\textbf {\bibinfo {volume} {453}},\ \bibinfo {pages} {1023} (\bibinfo {year} {2008})}\BibitemShut {NoStop}%
	\bibitem [{\citenamefont {Gonz{\'a}lez-Tudela}\ \emph {et~al.}(2024)\citenamefont {Gonz{\'a}lez-Tudela}, \citenamefont {Reiserer}, \citenamefont {Garc{\'\i}a-Ripoll},\ and\ \citenamefont {Garc{\'\i}a-Vidal}}]{Gonzalez-Tudela2024}%
	\BibitemOpen
	\bibfield  {author} {\bibinfo {author} {\bibfnamefont {A.}~\bibnamefont {Gonz{\'a}lez-Tudela}}, \bibinfo {author} {\bibfnamefont {A.}~\bibnamefont {Reiserer}}, \bibinfo {author} {\bibfnamefont {J.~J.}\ \bibnamefont {Garc{\'\i}a-Ripoll}},\ and\ \bibinfo {author} {\bibfnamefont {F.~J.}\ \bibnamefont {Garc{\'\i}a-Vidal}},\ }\bibfield  {title} {\bibinfo {title} {Light--matter interactions in quantum nanophotonic devices},\ }\href {https://doi.org/10.1038/s42254-023-00681-1} {\bibfield  {journal} {\bibinfo  {journal} {Nature Reviews Physics}\ }\textbf {\bibinfo {volume} {6}},\ \bibinfo {pages} {166} (\bibinfo {year} {2024})}\BibitemShut {NoStop}%
	\bibitem [{\citenamefont {Tey}\ \emph {et~al.}(2008)\citenamefont {Tey}, \citenamefont {Chen}, \citenamefont {Aljunid}, \citenamefont {Chng}, \citenamefont {Huber}, \citenamefont {Maslennikov},\ and\ \citenamefont {Kurtsiefer}}]{Tey2008}%
	\BibitemOpen
	\bibfield  {author} {\bibinfo {author} {\bibfnamefont {M.~K.}\ \bibnamefont {Tey}}, \bibinfo {author} {\bibfnamefont {Z.}~\bibnamefont {Chen}}, \bibinfo {author} {\bibfnamefont {S.~A.}\ \bibnamefont {Aljunid}}, \bibinfo {author} {\bibfnamefont {B.}~\bibnamefont {Chng}}, \bibinfo {author} {\bibfnamefont {F.}~\bibnamefont {Huber}}, \bibinfo {author} {\bibfnamefont {G.}~\bibnamefont {Maslennikov}},\ and\ \bibinfo {author} {\bibfnamefont {C.}~\bibnamefont {Kurtsiefer}},\ }\bibfield  {title} {\bibinfo {title} {Strong interaction between light and a single trapped atom without the need for a cavity},\ }\href {https://doi.org/10.1038/nphys1096} {\bibfield  {journal} {\bibinfo  {journal} {Nat. Phys.}\ }\textbf {\bibinfo {volume} {4}},\ \bibinfo {pages} {924} (\bibinfo {year} {2008})}\BibitemShut {NoStop}%
	\bibitem [{\citenamefont {Dicke}(1954)}]{dicke1954}%
	\BibitemOpen
	\bibfield  {author} {\bibinfo {author} {\bibfnamefont {R.~H.}\ \bibnamefont {Dicke}},\ }\bibfield  {title} {\bibinfo {title} {Coherence in spontaneous radiation processes},\ }\href {https://doi.org/10.1103/PhysRev.93.99} {\bibfield  {journal} {\bibinfo  {journal} {Phys. Rev.}\ }\textbf {\bibinfo {volume} {93}},\ \bibinfo {pages} {99} (\bibinfo {year} {1954})}\BibitemShut {NoStop}%
	\bibitem [{\citenamefont {Rui}\ \emph {et~al.}(2020)\citenamefont {Rui}, \citenamefont {Wei}, \citenamefont {Rubio-Abadal}, \citenamefont {Hollerith}, \citenamefont {Zeiher}, \citenamefont {Stamper-Kurn}, \citenamefont {Gross},\ and\ \citenamefont {Bloch}}]{Rui2020}%
	\BibitemOpen
	\bibfield  {author} {\bibinfo {author} {\bibfnamefont {J.}~\bibnamefont {Rui}}, \bibinfo {author} {\bibfnamefont {D.}~\bibnamefont {Wei}}, \bibinfo {author} {\bibfnamefont {A.}~\bibnamefont {Rubio-Abadal}}, \bibinfo {author} {\bibfnamefont {S.}~\bibnamefont {Hollerith}}, \bibinfo {author} {\bibfnamefont {J.}~\bibnamefont {Zeiher}}, \bibinfo {author} {\bibfnamefont {D.~M.}\ \bibnamefont {Stamper-Kurn}}, \bibinfo {author} {\bibfnamefont {C.}~\bibnamefont {Gross}},\ and\ \bibinfo {author} {\bibfnamefont {I.}~\bibnamefont {Bloch}},\ }\bibfield  {title} {\bibinfo {title} {A subradiant optical mirror formed by a single structured atomic layer},\ }\href {https://doi.org/10.1038/s41586-020-2463-x} {\bibfield  {journal} {\bibinfo  {journal} {Nature}\ }\textbf {\bibinfo {volume} {583}},\ \bibinfo {pages} {369} (\bibinfo {year} {2020})}\BibitemShut {NoStop}%
	\bibitem [{\citenamefont {Wilk}\ \emph {et~al.}(2007)\citenamefont {Wilk}, \citenamefont {Webster}, \citenamefont {Kuhn},\ and\ \citenamefont {Rempe}}]{wilk2007}%
	\BibitemOpen
	\bibfield  {author} {\bibinfo {author} {\bibfnamefont {T.}~\bibnamefont {Wilk}}, \bibinfo {author} {\bibfnamefont {S.~C.}\ \bibnamefont {Webster}}, \bibinfo {author} {\bibfnamefont {A.}~\bibnamefont {Kuhn}},\ and\ \bibinfo {author} {\bibfnamefont {G.}~\bibnamefont {Rempe}},\ }\bibfield  {title} {\bibinfo {title} {Single-atom single-photon quantum interface},\ }\href {https://doi.org/10.1126/science.1143835} {\bibfield  {journal} {\bibinfo  {journal} {Science}\ }\textbf {\bibinfo {volume} {317}},\ \bibinfo {pages} {488} (\bibinfo {year} {2007})}\BibitemShut {NoStop}%
	\bibitem [{\citenamefont {Arcari}\ \emph {et~al.}(2014)\citenamefont {Arcari}, \citenamefont {S{\"o}llner}, \citenamefont {Javadi}, \citenamefont {Lindskov~Hansen}, \citenamefont {Mahmoodian}, \citenamefont {Liu}, \citenamefont {Thyrrestrup}, \citenamefont {Lee}, \citenamefont {Song}, \citenamefont {Stobbe},\ and\ \citenamefont {Lodahl}}]{arcari2014}%
	\BibitemOpen
	\bibfield  {author} {\bibinfo {author} {\bibfnamefont {M.}~\bibnamefont {Arcari}}, \bibinfo {author} {\bibfnamefont {I.}~\bibnamefont {S{\"o}llner}}, \bibinfo {author} {\bibfnamefont {A.}~\bibnamefont {Javadi}}, \bibinfo {author} {\bibfnamefont {S.}~\bibnamefont {Lindskov~Hansen}}, \bibinfo {author} {\bibfnamefont {S.}~\bibnamefont {Mahmoodian}}, \bibinfo {author} {\bibfnamefont {J.}~\bibnamefont {Liu}}, \bibinfo {author} {\bibfnamefont {H.}~\bibnamefont {Thyrrestrup}}, \bibinfo {author} {\bibfnamefont {E.~H.}\ \bibnamefont {Lee}}, \bibinfo {author} {\bibfnamefont {J.~D.}\ \bibnamefont {Song}}, \bibinfo {author} {\bibfnamefont {S.}~\bibnamefont {Stobbe}},\ and\ \bibinfo {author} {\bibfnamefont {P.}~\bibnamefont {Lodahl}},\ }\bibfield  {title} {\bibinfo {title} {Near-unity coupling efficiency of a quantum emitter to a photonic crystal waveguide},\ }\href {https://doi.org/10.1103/PhysRevLett.113.093603} {\bibfield  {journal} {\bibinfo  {journal} {Phys. Rev. Lett.}\ }\textbf {\bibinfo {volume} {113}},\
		\bibinfo {pages} {093603} (\bibinfo {year} {2014})}\BibitemShut {NoStop}%
	\bibitem [{\citenamefont {Scarpelli}\ \emph {et~al.}(2019)\citenamefont {Scarpelli}, \citenamefont {Lang}, \citenamefont {Masia}, \citenamefont {Beggs}, \citenamefont {Muljarov}, \citenamefont {Young}, \citenamefont {Oulton}, \citenamefont {Kamp}, \citenamefont {H\"ofling}, \citenamefont {Schneider},\ and\ \citenamefont {Langbein}}]{Scarpelli2019}%
	\BibitemOpen
	\bibfield  {author} {\bibinfo {author} {\bibfnamefont {L.}~\bibnamefont {Scarpelli}}, \bibinfo {author} {\bibfnamefont {B.}~\bibnamefont {Lang}}, \bibinfo {author} {\bibfnamefont {F.}~\bibnamefont {Masia}}, \bibinfo {author} {\bibfnamefont {D.~M.}\ \bibnamefont {Beggs}}, \bibinfo {author} {\bibfnamefont {E.~A.}\ \bibnamefont {Muljarov}}, \bibinfo {author} {\bibfnamefont {A.~B.}\ \bibnamefont {Young}}, \bibinfo {author} {\bibfnamefont {R.}~\bibnamefont {Oulton}}, \bibinfo {author} {\bibfnamefont {M.}~\bibnamefont {Kamp}}, \bibinfo {author} {\bibfnamefont {S.}~\bibnamefont {H\"ofling}}, \bibinfo {author} {\bibfnamefont {C.}~\bibnamefont {Schneider}},\ and\ \bibinfo {author} {\bibfnamefont {W.}~\bibnamefont {Langbein}},\ }\bibfield  {title} {\bibinfo {title} {99
		}\textbf {\bibinfo {volume} {100}},\ \bibinfo {pages} {035311} (\bibinfo {year} {2019})}\BibitemShut {NoStop}%
	\bibitem [{\citenamefont {Shen}\ and\ \citenamefont {Fan}(2005)}]{shen2005}%
	\BibitemOpen
	\bibfield  {author} {\bibinfo {author} {\bibfnamefont {J.~T.}\ \bibnamefont {Shen}}\ and\ \bibinfo {author} {\bibfnamefont {S.}~\bibnamefont {Fan}},\ }\bibfield  {title} {\bibinfo {title} {Coherent photon transport from spontaneous emission in one-dimensional waveguides},\ }\href {https://doi.org/10.1364/OL.30.002001} {\bibfield  {journal} {\bibinfo  {journal} {Opt. Lett.}\ }\textbf {\bibinfo {volume} {30}},\ \bibinfo {pages} {2001} (\bibinfo {year} {2005})}\BibitemShut {NoStop}%
	\bibitem [{\citenamefont {Astafiev}\ \emph {et~al.}(2010)\citenamefont {Astafiev}, \citenamefont {Zagoskin}, \citenamefont {Abdumalikov}, \citenamefont {Pashkin}, \citenamefont {Yamamoto}, \citenamefont {Inomata}, \citenamefont {Nakamura},\ and\ \citenamefont {Tsai}}]{Astafiev2010}%
	\BibitemOpen
	\bibfield  {author} {\bibinfo {author} {\bibfnamefont {O.}~\bibnamefont {Astafiev}}, \bibinfo {author} {\bibfnamefont {A.~M.}\ \bibnamefont {Zagoskin}}, \bibinfo {author} {\bibfnamefont {A.~A.}\ \bibnamefont {Abdumalikov}}, \bibinfo {author} {\bibfnamefont {Y.~A.}\ \bibnamefont {Pashkin}}, \bibinfo {author} {\bibfnamefont {T.}~\bibnamefont {Yamamoto}}, \bibinfo {author} {\bibfnamefont {K.}~\bibnamefont {Inomata}}, \bibinfo {author} {\bibfnamefont {Y.}~\bibnamefont {Nakamura}},\ and\ \bibinfo {author} {\bibfnamefont {J.~S.}\ \bibnamefont {Tsai}},\ }\bibfield  {title} {\bibinfo {title} {Resonance fluorescence of a single artificial atom},\ }\href {https://doi.org/10.1126/science.1181918} {\bibfield  {journal} {\bibinfo  {journal} {Science}\ }\textbf {\bibinfo {volume} {327}},\ \bibinfo {pages} {840} (\bibinfo {year} {2010})}\BibitemShut {NoStop}%
	\bibitem [{\citenamefont {Chen}\ \emph {et~al.}(2011)\citenamefont {Chen}, \citenamefont {Wubs}, \citenamefont {M{\o}rk},\ and\ \citenamefont {Koenderink}}]{chen2011}%
	\BibitemOpen
	\bibfield  {author} {\bibinfo {author} {\bibfnamefont {Y.}~\bibnamefont {Chen}}, \bibinfo {author} {\bibfnamefont {M.}~\bibnamefont {Wubs}}, \bibinfo {author} {\bibfnamefont {J.}~\bibnamefont {M{\o}rk}},\ and\ \bibinfo {author} {\bibfnamefont {A.~F.}\ \bibnamefont {Koenderink}},\ }\bibfield  {title} {\bibinfo {title} {Coherent single-photon absorption by single emitters coupled to one-dimensional nanophotonic waveguides},\ }\href {https://doi.org/2011101116130300} {\bibfield  {journal} {\bibinfo  {journal} {New J. Phys.}\ }\textbf {\bibinfo {volume} {13}},\ \bibinfo {pages} {103010} (\bibinfo {year} {2011})}\BibitemShut {NoStop}%
	\bibitem [{\citenamefont {Zheng}\ \emph {et~al.}(2010)\citenamefont {Zheng}, \citenamefont {Gauthier},\ and\ \citenamefont {Baranger}}]{zheng2010}%
	\BibitemOpen
	\bibfield  {author} {\bibinfo {author} {\bibfnamefont {H.}~\bibnamefont {Zheng}}, \bibinfo {author} {\bibfnamefont {D.~J.}\ \bibnamefont {Gauthier}},\ and\ \bibinfo {author} {\bibfnamefont {H.~U.}\ \bibnamefont {Baranger}},\ }\bibfield  {title} {\bibinfo {title} {Waveguide qed: Many-body bound-state effects in coherent and fock-state scattering from a two-level system},\ }\href {https://doi.org/10.1103/PhysRevA.82.063816} {\bibfield  {journal} {\bibinfo  {journal} {Phys. Rev. A}\ }\textbf {\bibinfo {volume} {82}},\ \bibinfo {pages} {063816} (\bibinfo {year} {2010})}\BibitemShut {NoStop}%
	\bibitem [{\citenamefont {Douglas}\ \emph {et~al.}(2015)\citenamefont {Douglas}, \citenamefont {Habibian}, \citenamefont {Hung}, \citenamefont {Gorshkov}, \citenamefont {Kimble},\ and\ \citenamefont {Chang}}]{douglas2015}%
	\BibitemOpen
	\bibfield  {author} {\bibinfo {author} {\bibfnamefont {J.~S.}\ \bibnamefont {Douglas}}, \bibinfo {author} {\bibfnamefont {H.}~\bibnamefont {Habibian}}, \bibinfo {author} {\bibfnamefont {C.-L.}\ \bibnamefont {Hung}}, \bibinfo {author} {\bibfnamefont {A.~V.}\ \bibnamefont {Gorshkov}}, \bibinfo {author} {\bibfnamefont {H.~J.}\ \bibnamefont {Kimble}},\ and\ \bibinfo {author} {\bibfnamefont {D.~E.}\ \bibnamefont {Chang}},\ }\bibfield  {title} {\bibinfo {title} {Quantum many-body models with cold atoms coupled to photonic crystals},\ }\href {https://doi.org/10.1038/nphoton.2015.57} {\bibfield  {journal} {\bibinfo  {journal} {Nat. Photon.}\ }\textbf {\bibinfo {volume} {9}},\ \bibinfo {pages} {326} (\bibinfo {year} {2015})}\BibitemShut {NoStop}%
	\bibitem [{\citenamefont {Cheng}\ \emph {et~al.}(2017)\citenamefont {Cheng}, \citenamefont {Xu},\ and\ \citenamefont {Agarwal}}]{Cheng2017}%
	\BibitemOpen
	\bibfield  {author} {\bibinfo {author} {\bibfnamefont {M.-T.}\ \bibnamefont {Cheng}}, \bibinfo {author} {\bibfnamefont {J.}~\bibnamefont {Xu}},\ and\ \bibinfo {author} {\bibfnamefont {G.~S.}\ \bibnamefont {Agarwal}},\ }\bibfield  {title} {\bibinfo {title} {Waveguide transport mediated by strong coupling with atoms},\ }\href {https://doi.org/10.1103/PhysRevA.95.053807} {\bibfield  {journal} {\bibinfo  {journal} {Phys. Rev. A}\ }\textbf {\bibinfo {volume} {95}},\ \bibinfo {pages} {053807} (\bibinfo {year} {2017})}\BibitemShut {NoStop}%
	\bibitem [{\citenamefont {Poshakinskiy}\ and\ \citenamefont {Poddubny}(2021)}]{poshakinskiy2021}%
	\BibitemOpen
	\bibfield  {author} {\bibinfo {author} {\bibfnamefont {A.~V.}\ \bibnamefont {Poshakinskiy}}\ and\ \bibinfo {author} {\bibfnamefont {A.~N.}\ \bibnamefont {Poddubny}},\ }\bibfield  {title} {\bibinfo {title} {Dimerization of many-body subradiant states in waveguide quantum electrodynamics},\ }\href {https://doi.org/10.1103/PhysRevLett.127.173601} {\bibfield  {journal} {\bibinfo  {journal} {Phys. Rev. Lett.}\ }\textbf {\bibinfo {volume} {127}},\ \bibinfo {pages} {173601} (\bibinfo {year} {2021})}\BibitemShut {NoStop}%
	\bibitem [{\citenamefont {{R. Holzinger}}\ \emph {et~al.}(2022)\citenamefont {{R. Holzinger}}, \citenamefont {{R. Guti{\'e}rrez-J{\'a}uregui}}, \citenamefont {{T. H{\"o}nigl-Decrinis}}, \citenamefont {{G. Kirchmair}}, \citenamefont {{A. Asenjo-Garcia}},\ and\ \citenamefont {{H. Ritsch}}}]{r.holzinger2022}%
	\BibitemOpen
	\bibfield  {author} {\bibinfo {author} {\bibnamefont {{R. Holzinger}}}, \bibinfo {author} {\bibnamefont {{R. Guti{\'e}rrez-J{\'a}uregui}}}, \bibinfo {author} {\bibnamefont {{T. H{\"o}nigl-Decrinis}}}, \bibinfo {author} {\bibnamefont {{G. Kirchmair}}}, \bibinfo {author} {\bibnamefont {{A. Asenjo-Garcia}}},\ and\ \bibinfo {author} {\bibnamefont {{H. Ritsch}}},\ }\bibfield  {title} {\bibinfo {title} {Control of localized single- and many-body dark states in waveguide qed},\ }\href {https://doi.org/10.1103/PhysRevLett.129.253601} {\bibfield  {journal} {\bibinfo  {journal} {Phys. Rev. Lett.}\ }\textbf {\bibinfo {volume} {129}},\ \bibinfo {pages} {253601} (\bibinfo {year} {2022})}\BibitemShut {NoStop}%
	\bibitem [{\citenamefont {{Silvia Cardenas-Lopez}}\ \emph {et~al.}(2023)\citenamefont {{Silvia Cardenas-Lopez}}, \citenamefont {{Stuart J. Masson}}, \citenamefont {{Zoe Zager}},\ and\ \citenamefont {{Ana Asenjo-Garcia}}}]{silviacardenas-lopez2023}%
	\BibitemOpen
	\bibfield  {author} {\bibinfo {author} {\bibnamefont {{Silvia Cardenas-Lopez}}}, \bibinfo {author} {\bibnamefont {{Stuart J. Masson}}}, \bibinfo {author} {\bibnamefont {{Zoe Zager}}},\ and\ \bibinfo {author} {\bibnamefont {{Ana Asenjo-Garcia}}},\ }\bibfield  {title} {\bibinfo {title} {Many-body superradiance and dynamical mirror symmetry breaking in waveguide qed},\ }\href {https://doi.org/10.1103/PhysRevLett.131.033605} {\bibfield  {journal} {\bibinfo  {journal} {Phys. Rev. Lett.}\ }\textbf {\bibinfo {volume} {131}},\ \bibinfo {pages} {033605} (\bibinfo {year} {2023})}\BibitemShut {NoStop}%
	\bibitem [{\citenamefont {Fayard}\ \emph {et~al.}(2021)\citenamefont {Fayard}, \citenamefont {Henriet}, \citenamefont {{Asenjo-Garcia}},\ and\ \citenamefont {Chang}}]{fayard2021}%
	\BibitemOpen
	\bibfield  {author} {\bibinfo {author} {\bibfnamefont {N.}~\bibnamefont {Fayard}}, \bibinfo {author} {\bibfnamefont {L.}~\bibnamefont {Henriet}}, \bibinfo {author} {\bibfnamefont {A.}~\bibnamefont {{Asenjo-Garcia}}},\ and\ \bibinfo {author} {\bibfnamefont {D.~E.}\ \bibnamefont {Chang}},\ }\bibfield  {title} {\bibinfo {title} {Many-body localization in waveguide quantum electrodynamics},\ }\href {https://doi.org/10.1103/PhysRevResearch.3.033233} {\bibfield  {journal} {\bibinfo  {journal} {Phys. Rev. Res.}\ }\textbf {\bibinfo {volume} {3}},\ \bibinfo {pages} {033233} (\bibinfo {year} {2021})}\BibitemShut {NoStop}%
	\bibitem [{\citenamefont {Xing}\ \emph {et~al.}(2022)\citenamefont {Xing}, \citenamefont {Lu},\ and\ \citenamefont {Liao}}]{Xing2022}%
	\BibitemOpen
	\bibfield  {author} {\bibinfo {author} {\bibfnamefont {F.}~\bibnamefont {Xing}}, \bibinfo {author} {\bibfnamefont {Y.}~\bibnamefont {Lu}},\ and\ \bibinfo {author} {\bibfnamefont {Z.}~\bibnamefont {Liao}},\ }\bibfield  {title} {\bibinfo {title} {Quantum correlation propagation in a waveguide-qed system with long-range interaction},\ }\href {https://doi.org/10.1364/OE.462680} {\bibfield  {journal} {\bibinfo  {journal} {Opt. Express}\ }\textbf {\bibinfo {volume} {30}},\ \bibinfo {pages} {22963} (\bibinfo {year} {2022})}\BibitemShut {NoStop}%
	\bibitem [{\citenamefont {Sheremet}\ \emph {et~al.}(2023)\citenamefont {Sheremet}, \citenamefont {Petrov}, \citenamefont {Iorsh}, \citenamefont {Poshakinskiy},\ and\ \citenamefont {Poddubny}}]{sheremet2023}%
	\BibitemOpen
	\bibfield  {author} {\bibinfo {author} {\bibfnamefont {A.~S.}\ \bibnamefont {Sheremet}}, \bibinfo {author} {\bibfnamefont {M.~I.}\ \bibnamefont {Petrov}}, \bibinfo {author} {\bibfnamefont {I.~V.}\ \bibnamefont {Iorsh}}, \bibinfo {author} {\bibfnamefont {A.~V.}\ \bibnamefont {Poshakinskiy}},\ and\ \bibinfo {author} {\bibfnamefont {A.~N.}\ \bibnamefont {Poddubny}},\ }\bibfield  {title} {\bibinfo {title} {Waveguide quantum electrodynamics: Collective radiance and photon-photon correlations},\ }\href {https://doi.org/10.1103/RevModPhys.95.015002} {\bibfield  {journal} {\bibinfo  {journal} {Rev. Mod. Phys.}\ }\textbf {\bibinfo {volume} {95}},\ \bibinfo {pages} {015002} (\bibinfo {year} {2023})}\BibitemShut {NoStop}%
	\bibitem [{\citenamefont {Xing}\ \emph {et~al.}(2024{\natexlab{a}})\citenamefont {Xing}, \citenamefont {Wei},\ and\ \citenamefont {Liao}}]{Xing2024a}%
	\BibitemOpen
	\bibfield  {author} {\bibinfo {author} {\bibfnamefont {F.}~\bibnamefont {Xing}}, \bibinfo {author} {\bibfnamefont {Y.}~\bibnamefont {Wei}},\ and\ \bibinfo {author} {\bibfnamefont {Z.}~\bibnamefont {Liao}},\ }\bibfield  {title} {\bibinfo {title} {Quantum search in many-body interacting systems with long-range interactions},\ }\href {https://doi.org/10.1103/PhysRevA.109.052435} {\bibfield  {journal} {\bibinfo  {journal} {Phys. Rev. A}\ }\textbf {\bibinfo {volume} {109}},\ \bibinfo {pages} {052435} (\bibinfo {year} {2024}{\natexlab{a}})}\BibitemShut {NoStop}%
	\bibitem [{\citenamefont {Deutsch}\ \emph {et~al.}(1995)\citenamefont {Deutsch}, \citenamefont {Spreeuw}, \citenamefont {Rolston},\ and\ \citenamefont {Phillips}}]{Deutsch1995}%
	\BibitemOpen
	\bibfield  {author} {\bibinfo {author} {\bibfnamefont {I.~H.}\ \bibnamefont {Deutsch}}, \bibinfo {author} {\bibfnamefont {R.~J.~C.}\ \bibnamefont {Spreeuw}}, \bibinfo {author} {\bibfnamefont {S.~L.}\ \bibnamefont {Rolston}},\ and\ \bibinfo {author} {\bibfnamefont {W.~D.}\ \bibnamefont {Phillips}},\ }\bibfield  {title} {\bibinfo {title} {Photonic band gaps in optical lattices},\ }\href {https://doi.org/10.1103/PhysRevA.52.1394} {\bibfield  {journal} {\bibinfo  {journal} {Phys. Rev. A}\ }\textbf {\bibinfo {volume} {52}},\ \bibinfo {pages} {1394} (\bibinfo {year} {1995})}\BibitemShut {NoStop}%
	\bibitem [{\citenamefont {{T. S. Tsoi}}\ and\ \citenamefont {{C. K. Law}}(2008)}]{t.s.tsoi2008}%
	\BibitemOpen
	\bibfield  {author} {\bibinfo {author} {\bibnamefont {{T. S. Tsoi}}}\ and\ \bibinfo {author} {\bibnamefont {{C. K. Law}}},\ }\bibfield  {title} {\bibinfo {title} {Quantum interference effects of a single photon interacting with an atomic chain inside a one-dimensional waveguide},\ }\href {https://doi.org/10.1103/PhysRevA.78.063832} {\bibfield  {journal} {\bibinfo  {journal} {Phys. Rev. A}\ }\textbf {\bibinfo {volume} {78}},\ \bibinfo {pages} {063832} (\bibinfo {year} {2008})}\BibitemShut {NoStop}%
	\bibitem [{\citenamefont {Schilke}\ \emph {et~al.}(2011)\citenamefont {Schilke}, \citenamefont {Zimmermann}, \citenamefont {Courteille},\ and\ \citenamefont {Guerin}}]{Schilke2011}%
	\BibitemOpen
	\bibfield  {author} {\bibinfo {author} {\bibfnamefont {A.}~\bibnamefont {Schilke}}, \bibinfo {author} {\bibfnamefont {C.}~\bibnamefont {Zimmermann}}, \bibinfo {author} {\bibfnamefont {P.~W.}\ \bibnamefont {Courteille}},\ and\ \bibinfo {author} {\bibfnamefont {W.}~\bibnamefont {Guerin}},\ }\bibfield  {title} {\bibinfo {title} {Photonic band gaps in one-dimensionally ordered cold atomic vapors},\ }\href {https://doi.org/10.1103/PhysRevLett.106.223903} {\bibfield  {journal} {\bibinfo  {journal} {Phys. Rev. Lett.}\ }\textbf {\bibinfo {volume} {106}},\ \bibinfo {pages} {223903} (\bibinfo {year} {2011})}\BibitemShut {NoStop}%
	\bibitem [{\citenamefont {Liao}\ \emph {et~al.}(2015)\citenamefont {Liao}, \citenamefont {Zeng}, \citenamefont {Zhu},\ and\ \citenamefont {Zubairy}}]{liao2015}%
	\BibitemOpen
	\bibfield  {author} {\bibinfo {author} {\bibfnamefont {Z.}~\bibnamefont {Liao}}, \bibinfo {author} {\bibfnamefont {X.}~\bibnamefont {Zeng}}, \bibinfo {author} {\bibfnamefont {S.-Y.}\ \bibnamefont {Zhu}},\ and\ \bibinfo {author} {\bibfnamefont {M.~S.}\ \bibnamefont {Zubairy}},\ }\bibfield  {title} {\bibinfo {title} {Single-photon transport through an atomic chain coupled to a one-dimensional nanophotonic waveguide},\ }\href {https://doi.org/10.1103/PhysRevA.92.023806} {\bibfield  {journal} {\bibinfo  {journal} {Phys. Rev. A}\ }\textbf {\bibinfo {volume} {92}},\ \bibinfo {pages} {023806} (\bibinfo {year} {2015})}\BibitemShut {NoStop}%
	\bibitem [{\citenamefont {Corzo}\ \emph {et~al.}(2016)\citenamefont {Corzo}, \citenamefont {Gouraud}, \citenamefont {Chandra}, \citenamefont {Goban}, \citenamefont {Sheremet}, \citenamefont {Kupriyanov},\ and\ \citenamefont {Laurat}}]{corzo2016}%
	\BibitemOpen
	\bibfield  {author} {\bibinfo {author} {\bibfnamefont {N.~V.}\ \bibnamefont {Corzo}}, \bibinfo {author} {\bibfnamefont {B.}~\bibnamefont {Gouraud}}, \bibinfo {author} {\bibfnamefont {A.}~\bibnamefont {Chandra}}, \bibinfo {author} {\bibfnamefont {A.}~\bibnamefont {Goban}}, \bibinfo {author} {\bibfnamefont {A.~S.}\ \bibnamefont {Sheremet}}, \bibinfo {author} {\bibfnamefont {D.~V.}\ \bibnamefont {Kupriyanov}},\ and\ \bibinfo {author} {\bibfnamefont {J.}~\bibnamefont {Laurat}},\ }\bibfield  {title} {\bibinfo {title} {Large bragg reflection from one-dimensional chains of trapped atoms near a nanoscale waveguide},\ }\href {https://doi.org/10.1103/PhysRevLett.117.133603} {\bibfield  {journal} {\bibinfo  {journal} {Phys. Rev. Lett.}\ }\textbf {\bibinfo {volume} {117}},\ \bibinfo {pages} {133603} (\bibinfo {year} {2016})}\BibitemShut {NoStop}%
	\bibitem [{\citenamefont {S{\o}rensen}\ \emph {et~al.}(2016)\citenamefont {S{\o}rensen}, \citenamefont {B{\'e}guin}, \citenamefont {Kluge}, \citenamefont {Iakoupov}, \citenamefont {S{\o}rensen}, \citenamefont {M{\"u}ller}, \citenamefont {Polzik},\ and\ \citenamefont {Appel}}]{sorensen2016}%
	\BibitemOpen
	\bibfield  {author} {\bibinfo {author} {\bibfnamefont {H.~L.}\ \bibnamefont {S{\o}rensen}}, \bibinfo {author} {\bibfnamefont {J.-B.}\ \bibnamefont {B{\'e}guin}}, \bibinfo {author} {\bibfnamefont {K.~W.}\ \bibnamefont {Kluge}}, \bibinfo {author} {\bibfnamefont {I.}~\bibnamefont {Iakoupov}}, \bibinfo {author} {\bibfnamefont {A.~S.}\ \bibnamefont {S{\o}rensen}}, \bibinfo {author} {\bibfnamefont {J.~H.}\ \bibnamefont {M{\"u}ller}}, \bibinfo {author} {\bibfnamefont {E.~S.}\ \bibnamefont {Polzik}},\ and\ \bibinfo {author} {\bibfnamefont {J.}~\bibnamefont {Appel}},\ }\bibfield  {title} {\bibinfo {title} {Coherent backscattering of light off one-dimensional atomic strings},\ }\href {https://doi.org/10.1103/PhysRevLett.117.133604} {\bibfield  {journal} {\bibinfo  {journal} {Phys. Rev. Lett.}\ }\textbf {\bibinfo {volume} {117}},\ \bibinfo {pages} {133604} (\bibinfo {year} {2016})}\BibitemShut {NoStop}%
	\bibitem [{\citenamefont {Gonz\'alez-Tudela}\ \emph {et~al.}(2015)\citenamefont {Gonz\'alez-Tudela}, \citenamefont {Paulisch}, \citenamefont {Chang}, \citenamefont {Kimble},\ and\ \citenamefont {Cirac}}]{gonzalez-tudela2015}%
	\BibitemOpen
	\bibfield  {author} {\bibinfo {author} {\bibfnamefont {A.}~\bibnamefont {Gonz{\'a}lez-Tudela}}, \bibinfo {author} {\bibfnamefont {V.}~\bibnamefont {Paulisch}}, \bibinfo {author} {\bibfnamefont {D.~E.}\ \bibnamefont {Chang}}, \bibinfo {author} {\bibfnamefont {H.~J.}\ \bibnamefont {Kimble}},\ and\ \bibinfo {author} {\bibfnamefont {J.~I.}\ \bibnamefont {Cirac}},\ }\bibfield  {title} {\bibinfo {title} {Deterministic generation of arbitrary photonic states assisted by dissipation},\ }\href {https://doi.org/10.1103/PhysRevLett.115.163603} {\bibfield  {journal} {\bibinfo  {journal} {Phys. Rev. Lett.}\ }\textbf {\bibinfo {volume} {115}},\ \bibinfo {pages} {163603} (\bibinfo {year} {2015})}\BibitemShut {NoStop}%
	\bibitem [{\citenamefont {{Gonz{\'a}lez-Tudela}}\ \emph {et~al.}(2017)\citenamefont {{Gonz{\'a}lez-Tudela}}, \citenamefont {Paulisch}, \citenamefont {Kimble},\ and\ \citenamefont {Cirac}}]{gonzalez-tudela2017}%
	\BibitemOpen
	\bibfield  {author} {\bibinfo {author} {\bibfnamefont {A.}~\bibnamefont {{Gonz{\'a}lez-Tudela}}}, \bibinfo {author} {\bibfnamefont {V.}~\bibnamefont {Paulisch}}, \bibinfo {author} {\bibfnamefont {H.~J.}\ \bibnamefont {Kimble}},\ and\ \bibinfo {author} {\bibfnamefont {J.~I.}\ \bibnamefont {Cirac}},\ }\bibfield  {title} {\bibinfo {title} {Efficient multiphoton generation in waveguide quantum electrodynamics},\ }\href {https://doi.org/10.1103/PhysRevLett.118.213601} {\bibfield  {journal} {\bibinfo  {journal} {Phys. Rev. Lett.}\ }\textbf {\bibinfo {volume} {118}},\ \bibinfo {pages} {213601} (\bibinfo {year} {2017})}\BibitemShut {NoStop}%
	\bibitem [{\citenamefont {Xing}\ \emph {et~al.}(2024{\natexlab{b}})\citenamefont {Xing}, \citenamefont {Liao},\ and\ \citenamefont {Wang}}]{xing2024}%
	\BibitemOpen
	\bibfield  {author} {\bibinfo {author} {\bibfnamefont {F.}~\bibnamefont {Xing}}, \bibinfo {author} {\bibfnamefont {Z.}~\bibnamefont {Liao}},\ and\ \bibinfo {author} {\bibfnamefont {X.-h.}\ \bibnamefont {Wang}},\ }\bibfield  {title} {\bibinfo {title} {Deterministic generation of arbitrary $n$-photon states in a waveguide-qed system},\ }\href {https://doi.org/10.1103/PhysRevA.109.013718} {\bibfield  {journal} {\bibinfo  {journal} {Phys. Rev. A}\ }\textbf {\bibinfo {volume} {109}},\ \bibinfo {pages} {013718} (\bibinfo {year} {2024}{\natexlab{b}})}\BibitemShut {NoStop}%
	\bibitem [{\citenamefont {Paulisch}\ \emph {et~al.}(2016)\citenamefont {Paulisch}, \citenamefont {Kimble},\ and\ \citenamefont {Gonz谩lez-Tudela}}]{Paulisch2016}%
	\BibitemOpen
	\bibfield  {author} {\bibinfo {author} {\bibfnamefont {V.}~\bibnamefont {Paulisch}}, \bibinfo {author} {\bibfnamefont {H.~J.}\ \bibnamefont {Kimble}},\ and\ \bibinfo {author} {\bibfnamefont {A.}~\bibnamefont {Gonz{\'a}lez-Tudela}},\ }\bibfield  {title} {\bibinfo {title} {Universal quantum computation in waveguide qed using decoherence free subspaces},\ }\href {https://doi.org/10.1088/1367-2630/18/4/043041} {\bibfield  {journal} {\bibinfo  {journal} {New J. Phys.}\ }\textbf {\bibinfo {volume} {18}},\ \bibinfo {pages} {043041} (\bibinfo {year} {2016})}\BibitemShut {NoStop}%
	\bibitem [{\citenamefont {Liao}\ \emph {et~al.}(2017)\citenamefont {Liao}, \citenamefont {Al-Amri},\ and\ \citenamefont {Zubairy}}]{Liao2017OE}%
	\BibitemOpen
	\bibfield  {author} {\bibinfo {author} {\bibfnamefont {Z.}~\bibnamefont {Liao}}, \bibinfo {author} {\bibfnamefont {M.}~\bibnamefont {Al-Amri}},\ and\ \bibinfo {author} {\bibfnamefont {M.~S.}\ \bibnamefont {Zubairy}},\ }\bibfield  {title} {\bibinfo {title} {Measurement of deep-subwavelength emitter separation in a waveguide-qed system},\ }\href {https://doi.org/10.1364/OE.25.031997} {\bibfield  {journal} {\bibinfo  {journal} {Opt. Express}\ }\textbf {\bibinfo {volume} {25}},\ \bibinfo {pages} {31997} (\bibinfo {year} {2017})}\BibitemShut {NoStop}%
	\bibitem [{\citenamefont {Zhou}\ \emph {et~al.}(2008)\citenamefont {Zhou}, \citenamefont {Dong}, \citenamefont {Liu}, \citenamefont {Sun},\ and\ \citenamefont {Nori}}]{Zhou2008}%
	\BibitemOpen
	\bibfield  {author} {\bibinfo {author} {\bibfnamefont {L.}~\bibnamefont {Zhou}}, \bibinfo {author} {\bibfnamefont {H.}~\bibnamefont {Dong}}, \bibinfo {author} {\bibfnamefont {Y.-x.}\ \bibnamefont {Liu}}, \bibinfo {author} {\bibfnamefont {C.~P.}\ \bibnamefont {Sun}},\ and\ \bibinfo {author} {\bibfnamefont {F.}~\bibnamefont {Nori}},\ }\bibfield  {title} {\bibinfo {title} {Quantum supercavity with atomic mirrors},\ }\href {https://doi.org/10.1103/PhysRevA.78.063827} {\bibfield  {journal} {\bibinfo  {journal} {Phys. Rev. A}\ }\textbf {\bibinfo {volume} {78}},\ \bibinfo {pages} {063827} (\bibinfo {year} {2008})}\BibitemShut {NoStop}%
	\bibitem [{\citenamefont {Chang}\ \emph {et~al.}(2012)\citenamefont {Chang}, \citenamefont {Jiang}, \citenamefont {Gorshkov},\ and\ \citenamefont {Kimble}}]{chang2012}%
	\BibitemOpen
	\bibfield  {author} {\bibinfo {author} {\bibfnamefont {D.~E.}\ \bibnamefont {Chang}}, \bibinfo {author} {\bibfnamefont {L.}~\bibnamefont {Jiang}}, \bibinfo {author} {\bibfnamefont {A.~V.}\ \bibnamefont {Gorshkov}},\ and\ \bibinfo {author} {\bibfnamefont {H.~J.}\ \bibnamefont {Kimble}},\ }\bibfield  {title} {\bibinfo {title} {Cavity qed with atomic mirrors},\ }\href {https://doi.org/10.1088/1367-2630/14/6/063003} {\bibfield  {journal} {\bibinfo  {journal} {New J. Phys.}\ }\textbf {\bibinfo {volume} {14}},\ \bibinfo {pages} {063003} (\bibinfo {year} {2012})}\BibitemShut {NoStop}%
	\bibitem [{\citenamefont {Guimond}\ \emph {et~al.}(2016)\citenamefont {Guimond}, \citenamefont {Roulet}, \citenamefont {Le},\ and\ \citenamefont {Scarani}}]{Guimond2016}%
	\BibitemOpen
	\bibfield  {author} {\bibinfo {author} {\bibfnamefont {P.-O.}\ \bibnamefont {Guimond}}, \bibinfo {author} {\bibfnamefont {A.}~\bibnamefont {Roulet}}, \bibinfo {author} {\bibfnamefont {H.~N.}\ \bibnamefont {Le}},\ and\ \bibinfo {author} {\bibfnamefont {V.}~\bibnamefont {Scarani}},\ }\bibfield  {title} {\bibinfo {title} {Rabi oscillation in a quantum cavity: Markovian and non-markovian dynamics},\ }\href {https://doi.org/10.1103/PhysRevA.93.023808} {\bibfield  {journal} {\bibinfo  {journal} {Phys. Rev. A}\ }\textbf {\bibinfo {volume} {93}},\ \bibinfo {pages} {023808} (\bibinfo {year} {2016})}\BibitemShut {NoStop}%
	\bibitem [{\citenamefont {Mirhosseini}\ \emph {et~al.}(2019)\citenamefont {Mirhosseini}, \citenamefont {Kim}, \citenamefont {Zhang}, \citenamefont {Sipahigil}, \citenamefont {Dieterle}, \citenamefont {Keller}, \citenamefont {{Asenjo-Garcia}}, \citenamefont {Chang},\ and\ \citenamefont {Painter}}]{mirhosseini2019}%
	\BibitemOpen
	\bibfield  {author} {\bibinfo {author} {\bibfnamefont {M.}~\bibnamefont {Mirhosseini}}, \bibinfo {author} {\bibfnamefont {E.}~\bibnamefont {Kim}}, \bibinfo {author} {\bibfnamefont {X.}~\bibnamefont {Zhang}}, \bibinfo {author} {\bibfnamefont {A.}~\bibnamefont {Sipahigil}}, \bibinfo {author} {\bibfnamefont {P.~B.}\ \bibnamefont {Dieterle}}, \bibinfo {author} {\bibfnamefont {A.~J.}\ \bibnamefont {Keller}}, \bibinfo {author} {\bibfnamefont {A.}~\bibnamefont {{Asenjo-Garcia}}}, \bibinfo {author} {\bibfnamefont {D.~E.}\ \bibnamefont {Chang}},\ and\ \bibinfo {author} {\bibfnamefont {O.}~\bibnamefont {Painter}},\ }\bibfield  {title} {\bibinfo {title} {Cavity quantum electrodynamics with atom-like mirrors},\ }\href {https://doi.org/10.1038/s41586-019-1196-1} {\bibfield  {journal} {\bibinfo  {journal} {Nature}\ }\textbf {\bibinfo {volume} {569}},\ \bibinfo {pages} {692} (\bibinfo {year} {2019})}\BibitemShut {NoStop}%
	\bibitem [{\citenamefont {Zhou}\ \emph {et~al.}(2022)\citenamefont {Zhou}, \citenamefont {Liao},\ and\ \citenamefont {Zubairy}}]{Zhou2022}%
	\BibitemOpen
	\bibfield  {author} {\bibinfo {author} {\bibfnamefont {C.}~\bibnamefont {Zhou}}, \bibinfo {author} {\bibfnamefont {Z.}~\bibnamefont {Liao}},\ and\ \bibinfo {author} {\bibfnamefont {M.~S.}\ \bibnamefont {Zubairy}},\ }\bibfield  {title} {\bibinfo {title} {Decay of a single photon in a cavity with atomic mirrors},\ }\href {https://doi.org/10.1103/PhysRevA.105.033705} {\bibfield  {journal} {\bibinfo  {journal} {Phys. Rev. A}\ }\textbf {\bibinfo {volume} {105}},\ \bibinfo {pages} {033705} (\bibinfo {year} {2022})}\BibitemShut {NoStop}%
	\bibitem [{\citenamefont {Zhou}\ \emph {et~al.}(2023)\citenamefont {Zhou}, \citenamefont {Maleki}, \citenamefont {Liao},\ and\ \citenamefont {Zubairy}}]{Zhou2023}%
	\BibitemOpen
	\bibfield  {author} {\bibinfo {author} {\bibfnamefont {C.}~\bibnamefont {Zhou}}, \bibinfo {author} {\bibfnamefont {Y.}~\bibnamefont {Maleki}}, \bibinfo {author} {\bibfnamefont {Z.}~\bibnamefont {Liao}},\ and\ \bibinfo {author} {\bibfnamefont {M.~S.}\ \bibnamefont {Zubairy}},\ }\bibfield  {title} {\bibinfo {title} {Emission of an atom inside a one-dimensional atomic cavity},\ }\href {https://doi.org/10.1103/PhysRevA.108.013708} {\bibfield  {journal} {\bibinfo  {journal} {Phys. Rev. A}\ }\textbf {\bibinfo {volume} {108}},\ \bibinfo {pages} {013708} (\bibinfo {year} {2023})}\BibitemShut {NoStop}%
	\bibitem [{\citenamefont {Liao}\ \emph {et~al.}(2016{\natexlab{a}})\citenamefont {Liao}, \citenamefont {Nha},\ and\ \citenamefont {Zubairy}}]{liao2016a}%
	\BibitemOpen
	\bibfield  {author} {\bibinfo {author} {\bibfnamefont {Z.}~\bibnamefont {Liao}}, \bibinfo {author} {\bibfnamefont {H.}~\bibnamefont {Nha}},\ and\ \bibinfo {author} {\bibfnamefont {M.~S.}\ \bibnamefont {Zubairy}},\ }\bibfield  {title} {\bibinfo {title} {Single-photon frequency-comb generation in a one-dimensional waveguide coupled to two atomic arrays},\ }\href {https://doi.org/10.1103/PhysRevA.93.033851} {\bibfield  {journal} {\bibinfo  {journal} {Phys. Rev. A}\ }\textbf {\bibinfo {volume} {93}},\ \bibinfo {pages} {033851} (\bibinfo {year} {2016}{\natexlab{a}})}\BibitemShut {NoStop}%
	\bibitem [{\citenamefont {Wicht}\ \emph {et~al.}(1997)\citenamefont {Wicht}, \citenamefont {Danzmann}, \citenamefont {Fleischhauer}, \citenamefont {Scully}, \citenamefont {M{\"u}ller},\ and\ \citenamefont {Rinkleff}}]{wicht1997}%
	\BibitemOpen
	\bibfield  {author} {\bibinfo {author} {\bibfnamefont {A.}~\bibnamefont {Wicht}}, \bibinfo {author} {\bibfnamefont {K.}~\bibnamefont {Danzmann}}, \bibinfo {author} {\bibfnamefont {M.}~\bibnamefont {Fleischhauer}}, \bibinfo {author} {\bibfnamefont {M.}~\bibnamefont {Scully}}, \bibinfo {author} {\bibfnamefont {G.}~\bibnamefont {M{\"u}ller}},\ and\ \bibinfo {author} {\bibfnamefont {R.-H.}\ \bibnamefont {Rinkleff}},\ }\bibfield  {title} {\bibinfo {title} {White-light cavities, atomic phase coherence, and gravitational wave detectors},\ }\href {https://doi.org/10.1016/S0030-4018(96)00579-2} {\bibfield  {journal} {\bibinfo  {journal} {Opt. Commun.}\ }\textbf {\bibinfo {volume} {134}},\ \bibinfo {pages} {431} (\bibinfo {year} {1997})}\BibitemShut {NoStop}%
	\bibitem [{\citenamefont {{G. S. Pati}}\ \emph {et~al.}(2007)\citenamefont {{G. S. Pati}}, \citenamefont {{M. Salit}}, \citenamefont {{K. Salit}},\ and\ \citenamefont {{M. S. Shahriar}}}]{g.s.pati2007}%
	\BibitemOpen
	\bibfield  {author} {\bibinfo {author} {\bibnamefont {{G. S. Pati}}}, \bibinfo {author} {\bibnamefont {{M. Salit}}}, \bibinfo {author} {\bibnamefont {{K. Salit}}},\ and\ \bibinfo {author} {\bibnamefont {{M. S. Shahriar}}},\ }\bibfield  {title} {\bibinfo {title} {Demonstration of a tunable-bandwidth white-light interferometer using anomalous dispersion in atomic vapor},\ }\href {https://doi.org/10.1103/PhysRevLett.99.133601} {\bibfield  {journal} {\bibinfo  {journal} {Phys. Rev. Lett.}\ }\textbf {\bibinfo {volume} {99}},\ \bibinfo {pages} {133601} (\bibinfo {year} {2007})}\BibitemShut {NoStop}%
	\bibitem [{\citenamefont {{Yiqiu Ma}}\ \emph {et~al.}(2015)\citenamefont {{Yiqiu Ma}}, \citenamefont {{Haixing Miao}}, \citenamefont {{Chunnong Zhao}},\ and\ \citenamefont {{Yanbei Chen}}}]{yiqiuma2015}%
	\BibitemOpen
	\bibfield  {author} {\bibinfo {author} {\bibnamefont {{Yiqiu Ma}}}, \bibinfo {author} {\bibnamefont {{Haixing Miao}}}, \bibinfo {author} {\bibnamefont {{Chunnong Zhao}}},\ and\ \bibinfo {author} {\bibnamefont {{Yanbei Chen}}},\ }\bibfield  {title} {\bibinfo {title} {Quantum noise of a white-light cavity using a double-pumped gain medium},\ }\href {https://doi.org/10.1103/PhysRevA.92.023807} {\bibfield  {journal} {\bibinfo  {journal} {Phys. Rev. A}\ }\textbf {\bibinfo {volume} {92}},\ \bibinfo {pages} {023807} (\bibinfo {year} {2015})}\BibitemShut {NoStop}%
	\bibitem [{\citenamefont {Othman}\ \emph {et~al.}(2018)\citenamefont {Othman}, \citenamefont {Yevick},\ and\ \citenamefont {{Al-Amri}}}]{othman2018}%
	\BibitemOpen
	\bibfield  {author} {\bibinfo {author} {\bibfnamefont {A.}~\bibnamefont {Othman}}, \bibinfo {author} {\bibfnamefont {D.}~\bibnamefont {Yevick}},\ and\ \bibinfo {author} {\bibfnamefont {M.}~\bibnamefont {{Al-Amri}}},\ }\bibfield  {title} {\bibinfo {title} {Generation of three wide frequency bands within a single white-light cavity},\ }\href {https://doi.org/10.1103/PhysRevA.97.043816} {\bibfield  {journal} {\bibinfo  {journal} {Phys. Rev. A}\ }\textbf {\bibinfo {volume} {97}},\ \bibinfo {pages} {043816} (\bibinfo {year} {2018})}\BibitemShut {NoStop}%
	\bibitem [{\citenamefont {Yum}\ \emph {et~al.}(2013)\citenamefont {Yum}, \citenamefont {Scheuer}, \citenamefont {Salit}, \citenamefont {Hemmer},\ and\ \citenamefont {Shahriar}}]{yum2013a}%
	\BibitemOpen
	\bibfield  {author} {\bibinfo {author} {\bibfnamefont {H.~N.}\ \bibnamefont {Yum}}, \bibinfo {author} {\bibfnamefont {J.}~\bibnamefont {Scheuer}}, \bibinfo {author} {\bibfnamefont {M.}~\bibnamefont {Salit}}, \bibinfo {author} {\bibfnamefont {P.~R.}\ \bibnamefont {Hemmer}},\ and\ \bibinfo {author} {\bibfnamefont {M.~S.}\ \bibnamefont {Shahriar}},\ }\bibfield  {title} {\bibinfo {title} {Demonstration of white light cavity effect using stimulated brillouin scattering in a fiber loop},\ }\href {https://doi.org/10.1109/JLT.2013.2288326} {\bibfield  {journal} {\bibinfo  {journal} {J. Lightwave Technol.}\ }\textbf {\bibinfo {volume} {31}},\ \bibinfo {pages} {3865} (\bibinfo {year} {2013})}\BibitemShut {NoStop}%
	\bibitem [{\citenamefont {Shi}\ \emph {et~al.}(2013)\citenamefont {Shi}, \citenamefont {Harris}, \citenamefont {Fenollosa}, \citenamefont {Rodriguez}, \citenamefont {Lu}, \citenamefont {Korgel},\ and\ \citenamefont {Meseguer}}]{shi2013}%
	\BibitemOpen
	\bibfield  {author} {\bibinfo {author} {\bibfnamefont {L.}~\bibnamefont {Shi}}, \bibinfo {author} {\bibfnamefont {J.~T.}\ \bibnamefont {Harris}}, \bibinfo {author} {\bibfnamefont {R.}~\bibnamefont {Fenollosa}}, \bibinfo {author} {\bibfnamefont {I.}~\bibnamefont {Rodriguez}}, \bibinfo {author} {\bibfnamefont {X.}~\bibnamefont {Lu}}, \bibinfo {author} {\bibfnamefont {B.~A.}\ \bibnamefont {Korgel}},\ and\ \bibinfo {author} {\bibfnamefont {F.}~\bibnamefont {Meseguer}},\ }\bibfield  {title} {\bibinfo {title} {Monodisperse silicon nanocavities and photonic crystals with magnetic response in the optical region},\ }\href {https://doi.org/10.1038/ncomms2934} {\bibfield  {journal} {\bibinfo  {journal} {Nat. Commun.}\ }\textbf {\bibinfo {volume} {4}},\ \bibinfo {pages} {1904} (\bibinfo {year} {2013})}\BibitemShut {NoStop}%
	\bibitem [{\citenamefont {Moitra}\ \emph {et~al.}(2014)\citenamefont {Moitra}, \citenamefont {Slovick}, \citenamefont {Gang~Yu}, \citenamefont {Krishnamurthy},\ and\ \citenamefont {Valentine}}]{moitra2014}%
	\BibitemOpen
	\bibfield  {author} {\bibinfo {author} {\bibfnamefont {P.}~\bibnamefont {Moitra}}, \bibinfo {author} {\bibfnamefont {B.~A.}\ \bibnamefont {Slovick}}, \bibinfo {author} {\bibfnamefont {Z.}~\bibnamefont {Gang~Yu}}, \bibinfo {author} {\bibfnamefont {S.}~\bibnamefont {Krishnamurthy}},\ and\ \bibinfo {author} {\bibfnamefont {J.}~\bibnamefont {Valentine}},\ }\bibfield  {title} {\bibinfo {title} {Experimental demonstration of a broadband all-dielectric metamaterial perfect reflector},\ }\href {https://doi.org/10.1063/1.4873521} {\bibfield  {journal} {\bibinfo  {journal} {Appl. Phys. Lett.}\ }\textbf {\bibinfo {volume} {104}},\ \bibinfo {pages} {171102} (\bibinfo {year} {2014})}\BibitemShut {NoStop}%
	\bibitem [{\citenamefont {Bendickson}\ \emph {et~al.}(1996)\citenamefont {Bendickson}, \citenamefont {Dowling},\ and\ \citenamefont {Scalora}}]{Bendickson1996}%
	\BibitemOpen
	\bibfield  {author} {\bibinfo {author} {\bibfnamefont {J.~M.}\ \bibnamefont {Bendickson}}, \bibinfo {author} {\bibfnamefont {J.~P.}\ \bibnamefont {Dowling}},\ and\ \bibinfo {author} {\bibfnamefont {M.}~\bibnamefont {Scalora}},\ }\bibfield  {title} {\bibinfo {title} {Analytic expressions for the electromagnetic mode density in finite, one-dimensional, photonic band-gap structures},\ }\href {https://doi.org/10.1103/PhysRevE.53.4107} {\bibfield  {journal} {\bibinfo  {journal} {Phys. Rev. E}\ }\textbf {\bibinfo {volume} {53}},\ \bibinfo {pages} {4107} (\bibinfo {year} {1996})}\BibitemShut {NoStop}%
	\bibitem [{\citenamefont {Saglamyurek}\ \emph {et~al.}(2011)\citenamefont {Saglamyurek}, \citenamefont {Sinclair}, \citenamefont {Jin}, \citenamefont {Slater}, \citenamefont {Oblak}, \citenamefont {Bussi{\`e}res}, \citenamefont {George}, \citenamefont {Ricken}, \citenamefont {Sohler},\ and\ \citenamefont {Tittel}}]{Erhan2011}%
	\BibitemOpen
	\bibfield  {author} {\bibinfo {author} {\bibfnamefont {E.}~\bibnamefont {Saglamyurek}}, \bibinfo {author} {\bibfnamefont {N.}~\bibnamefont {Sinclair}}, \bibinfo {author} {\bibfnamefont {J.}~\bibnamefont {Jin}}, \bibinfo {author} {\bibfnamefont {J.~A.}\ \bibnamefont {Slater}}, \bibinfo {author} {\bibfnamefont {D.}~\bibnamefont {Oblak}}, \bibinfo {author} {\bibfnamefont {F.}~\bibnamefont {Bussi{\`e}res}}, \bibinfo {author} {\bibfnamefont {M.}~\bibnamefont {George}}, \bibinfo {author} {\bibfnamefont {R.}~\bibnamefont {Ricken}}, \bibinfo {author} {\bibfnamefont {W.}~\bibnamefont {Sohler}},\ and\ \bibinfo {author} {\bibfnamefont {W.}~\bibnamefont {Tittel}},\ }\bibfield  {title} {\bibinfo {title} {Broadband waveguide quantum memory for entangled photons},\ }\href {https://doi.org/10.1038/nature09719} {\bibfield  {journal} {\bibinfo  {journal} {Nature}\ }\textbf {\bibinfo {volume} {469}},\ \bibinfo {pages} {512} (\bibinfo {year} {2011})}\BibitemShut {NoStop}%
	\bibitem [{\citenamefont {Guo}\ \emph {et~al.}(2019)\citenamefont {Guo}, \citenamefont {Feng}, \citenamefont {Yang}, \citenamefont {Yu}, \citenamefont {Chen}, \citenamefont {Yuan},\ and\ \citenamefont {Zhang}}]{Guo2019}%
	\BibitemOpen
	\bibfield  {author} {\bibinfo {author} {\bibfnamefont {J.}~\bibnamefont {Guo}}, \bibinfo {author} {\bibfnamefont {X.}~\bibnamefont {Feng}}, \bibinfo {author} {\bibfnamefont {P.}~\bibnamefont {Yang}}, \bibinfo {author} {\bibfnamefont {Z.}~\bibnamefont {Yu}}, \bibinfo {author} {\bibfnamefont {L.~Q.}\ \bibnamefont {Chen}}, \bibinfo {author} {\bibfnamefont {C.-H.}\ \bibnamefont {Yuan}},\ and\ \bibinfo {author} {\bibfnamefont {W.}~\bibnamefont {Zhang}},\ }\bibfield  {title} {\bibinfo {title} {High-performance raman quantum memory with optimal control in room temperature atoms},\ }\href {https://doi.org/10.1038/s41467-018-08118-5} {\bibfield  {journal} {\bibinfo  {journal} {Nat. Commun.}\ }\textbf {\bibinfo {volume} {10}},\ \bibinfo {pages} {148} (\bibinfo {year} {2019})}\BibitemShut {NoStop}%
	\bibitem [{\citenamefont {Moiseev}\ \emph {et~al.}(2021)\citenamefont {Moiseev}, \citenamefont {Tashchilina}, \citenamefont {Moiseev},\ and\ \citenamefont {Sanders}}]{moiseev2021}%
	\BibitemOpen
	\bibfield  {author} {\bibinfo {author} {\bibfnamefont {E.~S.}\ \bibnamefont {Moiseev}}, \bibinfo {author} {\bibfnamefont {A.}~\bibnamefont {Tashchilina}}, \bibinfo {author} {\bibfnamefont {S.~A.}\ \bibnamefont {Moiseev}},\ and\ \bibinfo {author} {\bibfnamefont {B.~C.}\ \bibnamefont {Sanders}},\ }\bibfield  {title} {\bibinfo {title} {Broadband quantum memory in a cavity via zero spectral dispersion},\ }\href {https://doi.org/10.1088/1367-2630/ac0754} {\bibfield  {journal} {\bibinfo  {journal} {New J. Phys.}\ }\textbf {\bibinfo {volume} {23}},\ \bibinfo {pages} {063071} (\bibinfo {year} {2021})}\BibitemShut {NoStop}%
	\bibitem [{\citenamefont {Arnold}\ \emph {et~al.}(2022)\citenamefont {Arnold}, \citenamefont {Victora}, \citenamefont {Goggin},\ and\ \citenamefont {Kwiat}}]{Arnold2022}%
	\BibitemOpen
	\bibfield  {author} {\bibinfo {author} {\bibfnamefont {N.~T.}\ \bibnamefont {Arnold}}, \bibinfo {author} {\bibfnamefont {M.}~\bibnamefont {Victora}}, \bibinfo {author} {\bibfnamefont {M.~E.}\ \bibnamefont {Goggin}},\ and\ \bibinfo {author} {\bibfnamefont {P.~G.}\ \bibnamefont {Kwiat}},\ }\bibfield  {title} {\bibinfo {title} {Broad-bandwidth photonic quantum memory},\ }in\ \href {https://doi.org/10.1364/QUANTUM.2022.QM4B.3} {\emph {\bibinfo {booktitle} {Quantum 2.0 Conference and Exhibition}}}\ (\bibinfo  {publisher} {Optica Publishing Group},\ \bibinfo {year} {2022})\ p.\ \bibinfo {pages} {QM4B.3}\BibitemShut {NoStop}%
	\bibitem [{\citenamefont {Xu}\ \emph {et~al.}(2012)\citenamefont {Xu}, \citenamefont {{Al-Amri}}, \citenamefont {Yang}, \citenamefont {Zhu},\ and\ \citenamefont {Zubairy}}]{xu2012}%
	\BibitemOpen
	\bibfield  {author} {\bibinfo {author} {\bibfnamefont {J.}~\bibnamefont {Xu}}, \bibinfo {author} {\bibfnamefont {M.}~\bibnamefont {{Al-Amri}}}, \bibinfo {author} {\bibfnamefont {Y.}~\bibnamefont {Yang}}, \bibinfo {author} {\bibfnamefont {S.-Y.}\ \bibnamefont {Zhu}},\ and\ \bibinfo {author} {\bibfnamefont {M.~S.}\ \bibnamefont {Zubairy}},\ }\bibfield  {title} {\bibinfo {title} {Wide-band optical switch via white light cavity},\ }\href {https://doi.org/10.1103/PhysRevA.86.033828} {\bibfield  {journal} {\bibinfo  {journal} {Phys. Rev. A}\ }\textbf {\bibinfo {volume} {86}},\ \bibinfo {pages} {033828} (\bibinfo {year} {2012})}\BibitemShut {NoStop}%
	\bibitem [{\citenamefont {Xia}\ and\ \citenamefont {Twamley}(2013)}]{Xia2013}%
	\BibitemOpen
	\bibfield  {author} {\bibinfo {author} {\bibfnamefont {K.}~\bibnamefont {Xia}}\ and\ \bibinfo {author} {\bibfnamefont {J.}~\bibnamefont {Twamley}},\ }\bibfield  {title} {\bibinfo {title} {All-optical switching and router via the direct quantum control of coupling between cavity modes},\ }\href {https://doi.org/10.1103/PhysRevX.3.031013} {\bibfield  {journal} {\bibinfo  {journal} {Phys. Rev. X}\ }\textbf {\bibinfo {volume} {3}},\ \bibinfo {pages} {031013} (\bibinfo {year} {2013})}\BibitemShut {NoStop}%
	\bibitem [{\citenamefont {Zhu}\ \emph {et~al.}(2024)\citenamefont {Zhu}, \citenamefont {Wang}, \citenamefont {Huang}, \citenamefont {Wu}, \citenamefont {Zhao}, \citenamefont {Xiao}, \citenamefont {Wang}, \citenamefont {Davidson}, \citenamefont {Ou}, \citenamefont {Little},\ and\ \citenamefont {Chu}}]{Zhu2024}%
	\BibitemOpen
	\bibfield  {author} {\bibinfo {author} {\bibfnamefont {X.}~\bibnamefont {Zhu}}, \bibinfo {author} {\bibfnamefont {X.}~\bibnamefont {Wang}}, \bibinfo {author} {\bibfnamefont {Y.}~\bibnamefont {Huang}}, \bibinfo {author} {\bibfnamefont {L.}~\bibnamefont {Wu}}, \bibinfo {author} {\bibfnamefont {C.}~\bibnamefont {Zhao}}, \bibinfo {author} {\bibfnamefont {M.}~\bibnamefont {Xiao}}, \bibinfo {author} {\bibfnamefont {L.}~\bibnamefont {Wang}}, \bibinfo {author} {\bibfnamefont {R.}~\bibnamefont {Davidson}}, \bibinfo {author} {\bibfnamefont {Y.}~\bibnamefont {Ou}}, \bibinfo {author} {\bibfnamefont {B.~E.}\ \bibnamefont {Little}},\ and\ \bibinfo {author} {\bibfnamefont {S.~T.}\ \bibnamefont {Chu}},\ }\bibfield  {title} {\bibinfo {title} {Low-loss and polarization insensitive $32\times 4$ optical switch for roadm applications},\ }\href {https://doi.org/10.1038/s41377-024-01456-8} {\bibfield  {journal} {\bibinfo  {journal} {Light: Science \& Applications}\ }\textbf {\bibinfo {volume} {13}},\ \bibinfo {pages} {94} (\bibinfo
		{year} {2024})}\BibitemShut {NoStop}%
	\bibitem [{\citenamefont {Vengsarkar}\ \emph {et~al.}(1996)\citenamefont {Vengsarkar}, \citenamefont {Lemaire}, \citenamefont {Judkins}, \citenamefont {Bhatia}, \citenamefont {Erdogan},\ and\ \citenamefont {Sipe}}]{vengsarkar1996long}%
	\BibitemOpen
	\bibfield  {author} {\bibinfo {author} {\bibfnamefont {A.~M.}\ \bibnamefont {Vengsarkar}}, \bibinfo {author} {\bibfnamefont {P.~J.}\ \bibnamefont {Lemaire}}, \bibinfo {author} {\bibfnamefont {J.~B.}\ \bibnamefont {Judkins}}, \bibinfo {author} {\bibfnamefont {V.}~\bibnamefont {Bhatia}}, \bibinfo {author} {\bibfnamefont {T.}~\bibnamefont {Erdogan}},\ and\ \bibinfo {author} {\bibfnamefont {J.~E.}\ \bibnamefont {Sipe}},\ }\bibfield  {title} {\bibinfo {title} {Long-period fiber gratings as band-rejection filters},\ }\href@noop {} {\bibfield  {journal} {\bibinfo  {journal} {J. Lightwave Technol.}\ }\textbf {\bibinfo {volume} {14}},\ \bibinfo {pages} {58} (\bibinfo {year} {1996})}\BibitemShut {NoStop}%
	\bibitem [{\citenamefont {Lin}\ \emph {et~al.}(2023)\citenamefont {Lin}, \citenamefont {Lin}, \citenamefont {Xiao},\ and\ \citenamefont {Xiao}}]{Lin2023}%
	\BibitemOpen
	\bibfield  {author} {\bibinfo {author} {\bibfnamefont {H.}~\bibnamefont {Lin}}, \bibinfo {author} {\bibfnamefont {Y.}~\bibnamefont {Lin}}, \bibinfo {author} {\bibfnamefont {L.}~\bibnamefont {Xiao}},\ and\ \bibinfo {author} {\bibfnamefont {B.}~\bibnamefont {Xiao}},\ }\bibfield  {title} {\bibinfo {title} {Narrow-band rejection filter based on spoof surface plasmons polariton},\ }\href {https://doi.org/10.1007/s11082-023-04705-z} {\bibfield  {journal} {\bibinfo  {journal} {Optical and Quantum Electronics}\ }\textbf {\bibinfo {volume} {55}},\ \bibinfo {pages} {428} (\bibinfo {year} {2023})}\BibitemShut {NoStop}%
	\bibitem [{\citenamefont {Baumann}\ \emph {et~al.}(1996)\citenamefont {Baumann}, \citenamefont {Seifert}, \citenamefont {Nowak},\ and\ \citenamefont {Sauer}}]{baumann1996compact}%
	\BibitemOpen
	\bibfield  {author} {\bibinfo {author} {\bibfnamefont {I.}~\bibnamefont {Baumann}}, \bibinfo {author} {\bibfnamefont {J.}~\bibnamefont {Seifert}}, \bibinfo {author} {\bibfnamefont {W.}~\bibnamefont {Nowak}},\ and\ \bibinfo {author} {\bibfnamefont {M.}~\bibnamefont {Sauer}},\ }\bibfield  {title} {\bibinfo {title} {Compact all-fiber add-drop-multiplexer using fiber bragg gratings},\ }\href@noop {} {\bibfield  {journal} {\bibinfo  {journal} {IEEE Photonics Technology Letters}\ }\textbf {\bibinfo {volume} {8}},\ \bibinfo {pages} {1331} (\bibinfo {year} {1996})}\BibitemShut {NoStop}%
	\bibitem [{\citenamefont {Chen}\ \emph {et~al.}(2021)\citenamefont {Chen}, \citenamefont {Fontaine}, \citenamefont {Mazur}, \citenamefont {Ryf}, \citenamefont {Neilson}, \citenamefont {Song},\ and\ \citenamefont {Yan}}]{Chen2021}%
	\BibitemOpen
	\bibfield  {author} {\bibinfo {author} {\bibfnamefont {H.}~\bibnamefont {Chen}}, \bibinfo {author} {\bibfnamefont {N.~K.}\ \bibnamefont {Fontaine}}, \bibinfo {author} {\bibfnamefont {M.}~\bibnamefont {Mazur}}, \bibinfo {author} {\bibfnamefont {R.}~\bibnamefont {Ryf}}, \bibinfo {author} {\bibfnamefont {D.~T.}\ \bibnamefont {Neilson}}, \bibinfo {author} {\bibfnamefont {Q.}~\bibnamefont {Song}},\ and\ \bibinfo {author} {\bibfnamefont {Z.}~\bibnamefont {Yan}},\ }\bibfield  {title} {\bibinfo {title} {Wavelength selective switch components with high spectral resolution and compactness},\ }in\ \href {https://doi.org/10.1364/OFC.2021.W1A.1} {\emph {\bibinfo {booktitle} {Optical Fiber Communication Conference (OFC) 2021}}}\ (\bibinfo  {publisher} {Optica Publishing Group},\ \bibinfo {year} {2021})\ p.\ \bibinfo {pages} {W1A.1}\BibitemShut {NoStop}%
	\bibitem [{\citenamefont {Breglio}\ \emph {et~al.}(2006)\citenamefont {Breglio}, \citenamefont {Irace}, \citenamefont {Cusano},\ and\ \citenamefont {Cutolo}}]{breglio2006}%
	\BibitemOpen
	\bibfield  {author} {\bibinfo {author} {\bibfnamefont {G.}~\bibnamefont {Breglio}}, \bibinfo {author} {\bibfnamefont {A.}~\bibnamefont {Irace}}, \bibinfo {author} {\bibfnamefont {A.}~\bibnamefont {Cusano}},\ and\ \bibinfo {author} {\bibfnamefont {A.}~\bibnamefont {Cutolo}},\ }\bibfield  {title} {\bibinfo {title} {Chirped-pulsed frequency modulation (c-pfm) for fiber bragg grating sensors multiplexing},\ }\href {https://doi.org/10.1016/j.yofte.2005.06.002} {\bibfield  {journal} {\bibinfo  {journal} {Opt. Fiber Technol.}\ }\textbf {\bibinfo {volume} {12}},\ \bibinfo {pages} {71} (\bibinfo {year} {2006})}\BibitemShut {NoStop}%
	\bibitem [{\citenamefont {Lin}\ \emph {et~al.}(2024)\citenamefont {Lin}, \citenamefont {Wang}, \citenamefont {Lu}, \citenamefont {Hayle}, \citenamefont {Huang}, \citenamefont {Hsu}, \citenamefont {Chung}, \citenamefont {Bai}, \citenamefont {Okram},\ and\ \citenamefont {Lu}}]{Lin2024}%
	\BibitemOpen
	\bibfield  {author} {\bibinfo {author} {\bibfnamefont {H.-M.}\ \bibnamefont {Lin}}, \bibinfo {author} {\bibfnamefont {C.-P.}\ \bibnamefont {Wang}}, \bibinfo {author} {\bibfnamefont {H.-H.}\ \bibnamefont {Lu}}, \bibinfo {author} {\bibfnamefont {S.~T.}\ \bibnamefont {Hayle}}, \bibinfo {author} {\bibfnamefont {X.-H.}\ \bibnamefont {Huang}}, \bibinfo {author} {\bibfnamefont {W.-W.}\ \bibnamefont {Hsu}}, \bibinfo {author} {\bibfnamefont {Y.-C.}\ \bibnamefont {Chung}}, \bibinfo {author} {\bibfnamefont {Y.-Y.}\ \bibnamefont {Bai}}, \bibinfo {author} {\bibfnamefont {K.}~\bibnamefont {Okram}},\ and\ \bibinfo {author} {\bibfnamefont {J.-M.}\ \bibnamefont {Lu}},\ }\bibfield  {title} {\bibinfo {title} {Bidirectional wavelength-division-multiplexing fibre-free-space optical communications using polarisation multiplexing technique and tunable optical vestigial sideband filter},\ }\href {https://doi.org/10.1038/s44172-024-00277-2} {\bibfield  {journal} {\bibinfo  {journal} {Communs. Eng.}\ }\textbf {\bibinfo {volume}
			{3}},\ \bibinfo {pages} {128} (\bibinfo {year} {2024})}\BibitemShut {NoStop}%
	\bibitem [{\citenamefont {Liao}\ \emph {et~al.}(2016{\natexlab{b}})\citenamefont {Liao}, \citenamefont {Nha},\ and\ \citenamefont {Zubairy}}]{liao2016c}%
	\BibitemOpen
	\bibfield  {author} {\bibinfo {author} {\bibfnamefont {Z.}~\bibnamefont {Liao}}, \bibinfo {author} {\bibfnamefont {H.}~\bibnamefont {Nha}},\ and\ \bibinfo {author} {\bibfnamefont {M.~S.}\ \bibnamefont {Zubairy}},\ }\bibfield  {title} {\bibinfo {title} {Dynamical theory of single-photon transport in a one-dimensional waveguide coupled to identical and nonidentical emitters},\ }\href {https://doi.org/10.1103/PhysRevA.94.053842} {\bibfield  {journal} {\bibinfo  {journal} {Phys. Rev. A}\ }\textbf {\bibinfo {volume} {94}},\ \bibinfo {pages} {053842} (\bibinfo {year} {2016}{\natexlab{b}})}\BibitemShut {NoStop}%
	\bibitem [{\citenamefont {Liao}\ \emph {et~al.}(2020)\citenamefont {Liao}, \citenamefont {Lu},\ and\ \citenamefont {Zubairy}}]{liao2020}%
	\BibitemOpen
	\bibfield  {author} {\bibinfo {author} {\bibfnamefont {Z.}~\bibnamefont {Liao}}, \bibinfo {author} {\bibfnamefont {Y.}~\bibnamefont {Lu}},\ and\ \bibinfo {author} {\bibfnamefont {M.~S.}\ \bibnamefont {Zubairy}},\ }\bibfield  {title} {\bibinfo {title} {Multiphoton pulses interacting with multiple emitters in a one-dimensional waveguide},\ }\href {https://doi.org/10.1103/PhysRevA.102.053702} {\bibfield  {journal} {\bibinfo  {journal} {Phys. Rev. A}\ }\textbf {\bibinfo {volume} {102}},\ \bibinfo {pages} {053702} (\bibinfo {year} {2020})}\BibitemShut {NoStop}%
	\bibitem [{\citenamefont {{Asenjo-Garcia}}\ \emph {et~al.}(2017)\citenamefont {{Asenjo-Garcia}}, \citenamefont {{Moreno-Cardoner}}, \citenamefont {Albrecht}, \citenamefont {Kimble},\ and\ \citenamefont {Chang}}]{asenjo-garcia2017}%
	\BibitemOpen
	\bibfield  {author} {\bibinfo {author} {\bibfnamefont {A.}~\bibnamefont {{Asenjo-Garcia}}}, \bibinfo {author} {\bibfnamefont {M.}~\bibnamefont {{Moreno-Cardoner}}}, \bibinfo {author} {\bibfnamefont {A.}~\bibnamefont {Albrecht}}, \bibinfo {author} {\bibfnamefont {H.~J.}\ \bibnamefont {Kimble}},\ and\ \bibinfo {author} {\bibfnamefont {D.~E.}\ \bibnamefont {Chang}},\ }\bibfield  {title} {\bibinfo {title} {Exponential improvement in photon storage fidelities using subradiance and ``selective radiance'' in atomic arrays},\ }\href {https://doi.org/10.1103/PhysRevX.7.031024} {\bibfield  {journal} {\bibinfo  {journal} {Phys. Rev. X}\ }\textbf {\bibinfo {volume} {7}},\ \bibinfo {pages} {031024} (\bibinfo {year} {2017})}\BibitemShut {NoStop}%
	\bibitem [{\citenamefont {Albrecht}\ \emph {et~al.}(2019)\citenamefont {Albrecht}, \citenamefont {Henriet}, \citenamefont {{Asenjo-Garcia}}, \citenamefont {Dieterle}, \citenamefont {Painter},\ and\ \citenamefont {Chang}}]{albrecht2019}%
	\BibitemOpen
	\bibfield  {author} {\bibinfo {author} {\bibfnamefont {A.}~\bibnamefont {Albrecht}}, \bibinfo {author} {\bibfnamefont {L.}~\bibnamefont {Henriet}}, \bibinfo {author} {\bibfnamefont {A.}~\bibnamefont {{Asenjo-Garcia}}}, \bibinfo {author} {\bibfnamefont {P.~B.}\ \bibnamefont {Dieterle}}, \bibinfo {author} {\bibfnamefont {O.}~\bibnamefont {Painter}},\ and\ \bibinfo {author} {\bibfnamefont {D.~E.}\ \bibnamefont {Chang}},\ }\bibfield  {title} {\bibinfo {title} {Subradiant states of quantum bits coupled to a one-dimensional waveguide},\ }\href {https://doi.org/10.1088/1367-2630/ab0134} {\bibfield  {journal} {\bibinfo  {journal} {New J. Phys.}\ }\textbf {\bibinfo {volume} {21}},\ \bibinfo {pages} {025003} (\bibinfo {year} {2019})}\BibitemShut {NoStop}%
	\bibitem [{\citenamefont {Asenjo-Garcia}\ \emph {et~al.}(2017)\citenamefont {Asenjo-Garcia}, \citenamefont {Hood}, \citenamefont {Chang},\ and\ \citenamefont {Kimble}}]{PhysRevA.95.033818}%
	\BibitemOpen
	\bibfield  {author} {\bibinfo {author} {\bibfnamefont {A.}~\bibnamefont {Asenjo-Garcia}}, \bibinfo {author} {\bibfnamefont {J.~D.}\ \bibnamefont {Hood}}, \bibinfo {author} {\bibfnamefont {D.~E.}\ \bibnamefont {Chang}},\ and\ \bibinfo {author} {\bibfnamefont {H.~J.}\ \bibnamefont {Kimble}},\ }\bibfield  {title} {\bibinfo {title} {Atom-light interactions in quasi-one-dimensional nanostructures: A green's-function perspective},\ }\href {https://doi.org/10.1103/PhysRevA.95.033818} {\bibfield  {journal} {\bibinfo  {journal} {Phys. Rev. A}\ }\textbf {\bibinfo {volume} {95}},\ \bibinfo {pages} {033818} (\bibinfo {year} {2017})}\BibitemShut {NoStop}%
	\bibitem [{\citenamefont {Sheppard}(1995)}]{sheppard1995}%
	\BibitemOpen
	\bibfield  {author} {\bibinfo {author} {\bibfnamefont {C.~J.~R.}\ \bibnamefont {Sheppard}},\ }\bibfield  {title} {\bibinfo {title} {Approximate calculation of the reflection coefficient from a stratified medium},\ }\href {https://doi.org/10.1088/0963-9659/4/5/018} {\bibfield  {journal} {\bibinfo  {journal} {Pure and Applied Optics: Journal of the European Optical Society Part A}\ }\textbf {\bibinfo {volume} {4}},\ \bibinfo {pages} {665} (\bibinfo {year} {1995})}\BibitemShut {NoStop}%
	\bibitem [{\citenamefont {Jiang}\ \emph {et~al.}(2022)\citenamefont {Jiang}, \citenamefont {Xiao}, \citenamefont {Dong}, \citenamefont {Song},\ and\ \citenamefont {Wang}}]{jiang2022a}%
	\BibitemOpen
	\bibfield  {author} {\bibinfo {author} {\bibfnamefont {F.}~\bibnamefont {Jiang}}, \bibinfo {author} {\bibfnamefont {Z.}~\bibnamefont {Xiao}}, \bibinfo {author} {\bibfnamefont {M.}~\bibnamefont {Dong}}, \bibinfo {author} {\bibfnamefont {J.}~\bibnamefont {Song}},\ and\ \bibinfo {author} {\bibfnamefont {Y.}~\bibnamefont {Wang}},\ }\bibfield  {title} {\bibinfo {title} {Effect of distributed bragg reflectors on photoluminescence properties of ch3nh3pbi3 film},\ }\href {https://doi.org/10.1038/s41598-022-14991-4} {\bibfield  {journal} {\bibinfo  {journal} {Sci. Rep.}\ }\textbf {\bibinfo {volume} {12}},\ \bibinfo {pages} {10934} (\bibinfo {year} {2022})}\BibitemShut {NoStop}%
	\bibitem [{\citenamefont {Shen}\ and\ \citenamefont {Fan}(2007)}]{shen2007}%
	\BibitemOpen
	\bibfield  {author} {\bibinfo {author} {\bibfnamefont {J.-T.}\ \bibnamefont {Shen}}\ and\ \bibinfo {author} {\bibfnamefont {S.}~\bibnamefont {Fan}},\ }\bibfield  {title} {\bibinfo {title} {Strongly correlated two-photon transport in a one-dimensional waveguide coupled to a two-level system},\ }\href {https://doi.org/10.1103/PhysRevLett.98.153003} {\bibfield  {journal} {\bibinfo  {journal} {Phys. Rev. Lett.}\ }\textbf {\bibinfo {volume} {98}},\ \bibinfo {pages} {153003} (\bibinfo {year} {2007})}\BibitemShut {NoStop}%
	\bibitem [{\citenamefont {Roy}\ \emph {et~al.}(2017)\citenamefont {Roy}, \citenamefont {Wilson},\ and\ \citenamefont {Firstenberg}}]{roy2017}%
	\BibitemOpen
	\bibfield  {author} {\bibinfo {author} {\bibfnamefont {D.}~\bibnamefont {Roy}}, \bibinfo {author} {\bibfnamefont {C.~M.}\ \bibnamefont {Wilson}},\ and\ \bibinfo {author} {\bibfnamefont {O.}~\bibnamefont {Firstenberg}},\ }\bibfield  {title} {\bibinfo {title} {Colloquium: Strongly interacting photons in one-dimensional continuum},\ }\href {https://doi.org/10.1103/RevModPhys.89.021001} {\bibfield  {journal} {\bibinfo  {journal} {Rev. Mod. Phys.}\ }\textbf {\bibinfo {volume} {89}},\ \bibinfo {pages} {021001} (\bibinfo {year} {2017})}\BibitemShut {NoStop}%
	\bibitem [{\citenamefont {{Y. D. Chong}}\ \emph {et~al.}(2010)\citenamefont {{Y. D. Chong}}, \citenamefont {{Li Ge}}, \citenamefont {{Hui Cao}},\ and\ \citenamefont {{A. D. Stone}}}]{y.d.chong2010}%
	\BibitemOpen
	\bibfield  {author} {\bibinfo {author} {\bibnamefont {{Y. D. Chong}}}, \bibinfo {author} {\bibnamefont {{Li Ge}}}, \bibinfo {author} {\bibnamefont {{Hui Cao}}},\ and\ \bibinfo {author} {\bibnamefont {{A. D. Stone}}},\ }\bibfield  {title} {\bibinfo {title} {Coherent perfect absorbers: Time-reversed lasers},\ }\href {https://doi.org/10.1103/PhysRevLett.105.053901} {\bibfield  {journal} {\bibinfo  {journal} {Phys. Rev. Lett.}\ }\textbf {\bibinfo {volume} {105}},\ \bibinfo {pages} {053901} (\bibinfo {year} {2010})}\BibitemShut {NoStop}%
	\bibitem [{\citenamefont {{Sophocles J Orfanidis}}(2002)}]{sophoclesjorfanidis2002}%
	\BibitemOpen
	\bibfield  {author} {\bibinfo {author} {\bibnamefont {{Sophocles J Orfanidis}}},\ }\href@noop {} {\emph {\bibinfo {title} {Electromagnetic Waves and Antennas}}}\ (\bibinfo  {publisher} {Rutgers University New Brunswick, NJ},\ \bibinfo {year} {2002})\BibitemShut {NoStop}%
\end{thebibliography}
%

\end{document}